\documentclass[twocolumn,seceq]{jpsj3}

\def\runauthor{Online-Journal Subcommittee}

\usepackage{enumerate, graphics}

\newcommand{\bmath}[1]{\mbox{\boldmath $#1$}}

\title{Dynamics of the One-Dimensional 
Ising Model\\ without Detailed Balance Condition}

\author{Yuji SAKAI\thanks{E-mail address: 
yuji0920@huku.c.u-tokyo.ac.jp} and Koji HUKUSHIMA}
\inst{Department of Basic Science, 
Graduate School of Arts and Sciences,\\
The University of Tokyo, 3-8-1 Komaba, Meguro-ku, Tokyo 153-8902}

\abst{
We study an irreversible Markov chain Monte Carlo method based on a  
skew detailed balance condition for an one-dimensional Ising model. 
Dynamical behavior of the magnetization density is analyzed 
in order to understand the properties of this method. 
As a result, it is found theoretically that the relaxation time 
of the magnetization density is reduced by using some transition 
probabilities satisfying the skew detailed balance condition, 
in comparison to that with the corresponding transition probability 
with the detailed balance condition, 
and that one of the transition probabilities changes the dynamical 
critical exponent even with a local spin update.
}

\recdate{\today}

\kword{Markov chain Monte Carlo, Ising model, 
Glauber dynamics, detailed balance condition}

\begin{document}
\maketitle

\section{Introduction}

Markov chain Monte Carlo (MCMC) methods have been widely used for
sampling from a high-dimensional probability distribution and 
for estimating expectation values under the distribution. 
Since Metropolis et al. 
have proposed a MCMC method as a simulation tool for studying 
liquid~\cite{Metropolis}, it has been applied to various problems
in physics as well as other research fields.
The conventional MCMC methods, such as Metropolis-Hastings 
algorithm~\cite{Hastings} and heat-bath algorithm, have developed 
within the framework of the detailed balance condition (DBC),
which ensures the existence of stationary distribution in a Markov chain. 
It is, however, not always necessary to make the MCMC method 
work correctly.
MCMC methods which are not based on  DBC have been discussed  
for improving the performance and, 
in fact, some MCMC methods without DBC have recently been 
proposed~\cite{Suwa-Todo,Fernandes,Turitsyn}.

In the MCMC method, the convergence to the target distribution
is guaranteed by using an irreducible and aperiodic Markov chain,
characterized by a transition matrix. 
It is known that the rate of convergence of the distribution depends on 
the second largest eigenvalue of the transition matrix of the Markov chain.
Because all the eigenvalues are real if the transition matrix satisfies DBC, 
an estimator converges to its asymptotic value exponentially. 
On the other hand, if the transition matrix does not satisfy DBC, 
some eigenvalues could be complex in principle 
and then the estimator behaves like a damped oscillation.   
Such dynamics may affect the efficiency of performance of MCMC methods. 

Let us consider the efficiency of the MCMC method.
When one estimates the expectation value $\left<\mathcal{O}\right>$ 
of an observable $\mathcal{O}$ precisely, Monte Carlo steps $M$ 
should be sufficiently large, depending on an actual algorithm and 
a probabilistic model to be studied, so that the empirical distribution
converges to the target distribution. 
Then, a MCMC method is considered to be efficient if an estimator rapidly 
converges to an exact expectation value and the variance of the
estimator is sufficient small. 
The variance of the estimator is often enlarged by the correlation between
samples in the Markov chain, which inevitably appears 
in return for overcoming the curse of dimensionality in MCMC methods. 
The correlation between samples is evaluated from  
an integrated autocorrelation time $\tau_{\rm int,\mathcal{O}}$
defined by the autocorrelation function $C(t;\mathcal{O})$~\cite{Landau}:
	\begin{equation}
	\tau_{\rm int,\mathcal{O}}=
	\sum_{t=1}^{\infty}C(t;\mathcal{O})
	=\sum_{t=1}^{\infty}
	\frac{\left<\mathcal{O}_{i+t}\mathcal{O}_i\right>
	-\left<\mathcal{O}_i\right>^2}
	{\left<\mathcal{O}_i^2\right>
	-\left<\mathcal{O}_i\right>^2},
	\end{equation} 
where $\mathcal{O}_i$ is the value of $\mathcal{O}$ 
at $i$-th Monte Carlo step (MCS)
and $C(t;\mathcal{O})$ does not depend on $i$ 
after the system reaches equilibrium. 
The effective variance of the estimator is given as 
$\sigma^2_{\rm eff, \mathcal{O}}\simeq(1+2\tau_{\rm int,\mathcal{O}}) 
\sigma^2_{\mathcal{O}}$, 
where $\sigma^2_{\mathcal{O}}$ is the variance 
in the case of independent sampling.
The correlation time $\tau_{\rm int,\mathcal{O}}$ increases with
increase of the correlation between samples. 
Consequently, the number of effective samples decreases as
$M_{\rm eff}\simeq M/(1+2\tau_{\rm int,\mathcal{O}})$.
Thus, the efficient MCMC method requires the reduction of
the variance or the correlation time.  
 
In the Markov chain with DBC, 
Peskun's theorem provides us a guiding principle 
for constructing an efficient MCMC method~\cite{Peskun}.
According to the theorem, the asymptotic variance of any observable 
is reduced by decreasing the rejection rate of the Markov chain. 
From this point of view, it turns out that the Metropolis 
transition probability is more efficient than 
that used in the heat-bath method. 
However, this argument relies on DBC and no guiding principle has been
established in the case of the MCMC methods without DBC. 
It is therefore of great concern how MCMC methods are constructed without
the use of DBC and how the methods improve the efficiency of performance. 
For instance, Suwa and Todo have proposed 
a MCMC method without DBC~\cite{Suwa-Todo}, 
which brings several times of reduction in the correlation time of the
Potts model, 
in comparison to the corresponding method with DBC. 
Turitsyn et al.~\cite{Turitsyn} and Fernandes and
Weigel\cite{Fernandes} have also proposed other MCMC methods without DBC
and have suggested from numerical simulations that the dynamical
critical exponent may be changed in a mean-field Ising model. 
Although these numerical studies encourage to use the MCMC methods
without DBC, it is not well understood theoretically how 
these methods affect to the dynamics, 
particularly in statistical-mechanical models. 

In this paper, we discuss the dynamics in the  Markov chain without DBC in an
one-dimensional kinetic Ising model, which is exactly solved in the
Glauber dynamics with DBC~\cite{Glauber}.  
By solving time evolution of the order parameter for the model, 
it is found that some transition matrices yield the reduction of the
relaxation time which significantly depends on the choice of the
transition probability. 

The paper is organized as follows. In \S 2,  
static and dynamic properties of the one-dimensional 
Ising model are surveyed. 
In \S 3, an irreversible Markov chain based on the 
skew detailed balance condition is constructed for the Ising model and 
dynamical behavior of the magnetization density is analyzed in the
Markov chain.  
An outline of our MCMC simulation is described and their results 
are presented in \S 4.
Finally,  summary and discussions are given in \S 5.

\section{Reversible Glauber Dynamics}

Glauber dynamics is a Markov chain of configuration of an Ising model.
In this section, we survey the Glauber dynamics of an
one-dimensional Ising model with DBC~\cite{Glauber}
in order to fix our notation.

\subsection{Ising model in one dimension}

We study the one-dimensional Ising model
with no external magnetic field. 
A state of the Ising model is denoted by a vector 
$\bmath{\sigma}=(\sigma_1,\cdots,\sigma_N)$
with $\sigma_j=\pm 1$ being an Ising variable 
defined on $j$-th site.
The Hamiltonian is defined as 
	\begin{equation}
	\mathcal{H}(\bmath{\sigma})=
	-J\sum_{j=1}^{N}\sigma_j\sigma_{j+1}, 
	\label{eq:H}
	\end{equation}
where a periodic boundary condition is imposed as 
$\sigma_{N+1}=\sigma_1$ and 
$J$ is the exchange interaction constant. 
For a given inverse temperature $\beta$, in the units where
Boltzmann constant is 1, the equilibrium distribution 
$\pi(\bmath{\sigma})$ for finding a state $\bmath{\sigma}$ 
is proportional to the Boltzmann factor
$\exp\left(-\beta\mathcal{H}(\bmath{\sigma})\right)$.

An expectation of an 
observable $A=A(\bmath{\sigma})$ in equilibrium 
is expressed as 
	\begin{equation}
	\left<A\right>_{\rm eq}:=
	\sum_{\bmath{\sigma}}A(\bmath{\sigma})\pi(\bmath{\sigma}),
	\label{eq:expectation_eq}
	\end{equation}
where $\sum_{\bmath{\sigma}}$ denotes the summation over
$2^N$ spin configurations. The order parameter of the Ising model is 
magnetization density which is given in our notation by the expectation 
$\left<m\right>_{\rm eq}$ of the observable 
$m(\bmath{\sigma})=\frac1N\sum_j\sigma_j$.
It is well known that no spontaneous magnetization emerges 
at any finite temperature and the phase transition does not occur
in the one-dimensional Ising model, that is  
	\begin{equation}
	\left<m\right>_{\rm eq}=0,
	\label{eq:magnetization_density}
	\end{equation}
for all $N$ and $\beta>0$ in this model.
The correlation length $\xi(\beta)$ is given by 
	\begin{equation}
	\xi^{-1}(\beta)=-\log({\rm tanh}\beta J),
	\label{eq:correlation_length}
	\end{equation}
which diverges as the temperature goes to zero.

\subsection{Master equation and detailed balance condition}

Since the Ising model has no intrinsic dynamics induced by Hamiltonian, 
a stochastic dynamics introduced by Glauber~\cite{Glauber} has been used 
in the study of dynamics of the Ising models. 
The stochastic dynamics is a Markov chain of the state $\bmath{\sigma}$, 
which is described by a master equation.

We define $F_j$ be a spin-flip operator on $j$-th site: 
$F_j\bmath{\sigma}$ is the state that $j$-th spin is flipped 
from $\bmath{\sigma}$ with the others fixed. The Markov chain is 
characterized by a transition probability $w_j(\bmath{\sigma})$ 
per unit time from $\bmath{\sigma}$ to $F_j\bmath{\sigma}$. 
Let $p(\bmath{\sigma},t)$ be a probability distribution for 
finding the spin state $\bmath{\sigma}$ at time $t$.  
Then, the master equation is written as follows:
	\begin{eqnarray}
	\frac{d}{dt}p(\bmath{\sigma},t)
	&\!\!\!=\!\!\!&
	-\left[\sum_j w_j(\bmath{\sigma})\right]
	p(\bmath{\sigma},t)
	\notag \\ &&
	+\sum_j w_j(F_j\bmath{\sigma})p(F_j\bmath{\sigma},t),
	\label{eq:master}
	\end{eqnarray}
where the first and second terms in the right hand side are 
outgoing and incoming probability, respectively. 

For the master equation in Eq.~(\ref{eq:master}), 
the necessary and sufficient condition that 
$p(\bmath{\sigma},t)$ converges to the equilibrium distribution 
$\pi(\bmath{\sigma})$ as $t\to\infty$ is that the transition 
probability $w_j(\bmath{\sigma})$ satisfies a balance condition (BC):
	\begin{equation}
	\sum_j w_j(\bmath{\sigma})\pi(\bmath{\sigma})=
	\sum_j w_j(F_j\bmath{\sigma})\pi(F_j\bmath{\sigma}).
	\label{eq:BC}
	\end{equation}
In practice, DBC is widely used for a sufficient condition of BC: 
	\begin{equation}
	w_j(\bmath{\sigma})\pi(\bmath{\sigma})=
	w_j(F_j\bmath{\sigma})\pi(F_j\bmath{\sigma}).
	\label{eq:DBC}	
	\end{equation}
This condition is also called reversibility 
in the field of statistical science. 
The sequence of states generated by the transition probability 
with DBC is called reversible Markov chain.
While BC means the total balance of 
the stochastic flow in the state space,   
DBC requires a local balance of the stochastic flow 
between each state $\bmath{\sigma}$ and $F_j\bmath{\sigma}$. 
Imposing DBC, some transition probabilities can be determined explicitly 
and have been used in MCMC simulations.
For instance, Glauber's transition probability is given for the one-dimensional 
Ising model as 
	\begin{equation}
	w_j(\bmath{\sigma})=
	\frac12\alpha\left[1-\frac12\gamma\sigma_j
	(\sigma_{j-1}+\sigma_{j+1})\right], 
	\label{eq:w_j}
	\end{equation}
where $\alpha$ is a time constant and $\gamma={\rm tanh}2\beta J$. 
This is equivalent to the heat-bath algorithm in MCMC method~\cite{Landau}.

\subsection{Time evolution of magnetization density}

An expectation of an observable $A$ at time $t$ is denoted by
	\begin{equation}
	\left<A(t)\right>:=
	\sum_{\bmath{\sigma}}A(\bmath{\sigma})p(\bmath{\sigma},t). 
	\label{eq:expectation_t}
	\end{equation}
From the master equation in Eq.~(\ref{eq:master}), time evolution of the 
magnetization density is reduced to~\cite{Glauber}
	\begin{equation}
	\frac1\alpha\frac{d}{dt}\left<m(t)\right>=
	-(1-\gamma)\left<m(t)\right>. 
	\label{eq:time evolution of M}
	\end{equation}
Then, we have
	\begin{equation}
	\left<m(t)\right>=\left<m(0)\right>\exp[-\alpha(1-\gamma)t].
	\label{eq:solution of M}
	\end{equation}
This indicates that the 
magnetization density converges exponentially in time 
to the equilibrium value in Eq.~(\ref{eq:magnetization_density}).
The relaxation time of the magnetization density which reflects 
the rate of convergence is defined as
	\begin{equation}
	\tau:=\int_0^{\infty}\!\!\!dt
	\frac{\left<m(t)\right>-\left<m\right>_{\rm eq}}
	{\left<m(0)\right>-\left<m\right>_{\rm eq}}. 
	\label{eq:relaxation_time}
	\end{equation}
Using the solution of Eq.~(\ref{eq:solution of M}), the relaxation time 
of this system is obtained as 
	\begin{equation}
	\tau=\frac1{\alpha(1-\gamma)}, 
	\label{eq:relax}
	\end{equation}
which means that the convergence rate is slower with 
decreasing temperature and eventually diverges at zero temperature $\gamma=1$.
This is due to dynamical slowing down induced by the zero-temperature 
transition. For $\gamma\to 1$, one finds that $\tau\sim\xi^z$ using 
Eq.~(\ref{eq:correlation_length}) with the dynamical critical exponent $z=2$.

\subsection{Integrated autocorrelation time}

We also discuss an integrated autocorrelation time 
of the magnetization density. 
Let $p(\bmath{\sigma},t+t_{\omega}|\bmath{\sigma}', t_{\omega})$ 
be a conditional probability for finding a state 
$\bmath{\sigma}$ at elapsed time $t$ after a state $\bmath{\sigma}'$
at waiting time $t_{\omega}$. Then, 
an autocorrelation function of an observable $A$ in equilibrium is defined as
	\begin{equation}
	C_{\rm eq}(t;A):=\frac{\left<A_{\rm eq}A(t)\right>
	-\left<A\right>_{\rm eq}^2}
	{\left<A^2\right>_{\rm eq}-\left<A\right>_{\rm eq}^2}, 
	\label{eq:autocorrelation}
	\end{equation}
where
	\begin{equation}
	\left<A_{\rm eq}A(t)\right>:=
	\sum_{\bmath{\sigma},\bmath{\sigma}'}
	A(\bmath{\sigma}')\pi(\bmath{\sigma}')
	A(\bmath{\sigma})p(\bmath{\sigma},t+t_{\omega}
	|\bmath{\sigma}',t_{\omega}), 
	\label{eq:A_eqA(t)}
	\end{equation}
where $t_{\omega}$ being a sufficient long time.
In the case of the magnetization density, 
substituting Eq.~(\ref{eq:solution of M}) we have
	\begin{equation}
	\left<m_{\rm eq}m(t)\right>=\left<m^2\right>_{\rm eq}
	\exp[-\alpha(1-\gamma)t].
	\end{equation}
Combining with Eq.~(\ref{eq:magnetization_density}), we obtain
	\begin{equation}
	C_{\rm eq}(t;m)=\exp[-\alpha(1-\gamma)t]. 
	\end{equation}
Thus, the integrated autocorrelation time of the magnetization 
density is obtained as 
	\begin{equation}
	\tau_{{\rm int},m}:=\int_0^{\infty}\!\!\!dt~C_{\rm eq}(t;m)
	=\frac1{\alpha(1-\gamma)},
	\label{eq:autocorrelation_time}
	\end{equation}
which is identical to the relaxation time of the magnetization density. 
This indicates that 
the correlation between states in the Markov chain increases with temperature 
decreasing and consequently the effective variance of the magnetization density 
is enlarged.

\section{Irreversible Glauber Dynamics}

A stochastic process whose transition probability does not 
satisfy DBC is called irreversible Markov chain. 
In this section, we construct a Markov chain for the one-dimensional 
Ising model on a basis of skew detailed balance condition 
(SDBC)~\cite{Turitsyn}
and study time evolution of the magnetization density under SDBC.

\subsection{Master equation and skew detailed balance condition}

According to the method of Turitsyn et al.~\cite{Turitsyn}, 
we introduce another Ising spin 
$\varepsilon=\pm 1$ in addition to the original spin configurations. 
The enlarged state of the system is denoted by 
$X:=(\bmath{\sigma},\varepsilon)\in\left\{-1,+1\right\}^{N+1}$.
We consider a single spin-flip update for the whole spin 
including $\varepsilon$ as an elementary Markov process.  
Let $p(\bmath{\sigma},\varepsilon,t)$
be a probability distribution which we find a state 
$(\bmath{\sigma},\varepsilon)$ at time $t$. 
The master equation of the system is given as follows:
	\begin{eqnarray}
	\frac{d}{dt}p(\bmath{\sigma},\varepsilon,t)
	&\!\!\!=\!\!\!&
	-\left[\sum_j w_j(\bmath{\sigma},\varepsilon)\right]
	p(\bmath{\sigma},\varepsilon,t)
	\notag \\ &&
	+\sum_j w_j(F_j\bmath{\sigma},\varepsilon)
	p(F_j\bmath{\sigma},\varepsilon,t)
	\notag \\ &&
	-\lambda(\bmath{\sigma},\varepsilon)
	p(\bmath{\sigma},\varepsilon,t)
	\notag \\ &&
	+\lambda(\bmath{\sigma},-\varepsilon)
	p(\bmath{\sigma},-\varepsilon,t), 
	\label{eq:Smaster}
	\end{eqnarray}
where $w_j(\bmath{\sigma},\varepsilon)$ is a transition probability 
par unit time from a state $(\bmath{\sigma},\varepsilon)$ 
to $(F_j\bmath{\sigma},\varepsilon)$ and 
$\lambda(\bmath{\sigma},\varepsilon)$ is that 
from a state $(\bmath{\sigma},\varepsilon)$ 
to $(\bmath{\sigma},-\varepsilon)$.
We assume that an equilibrium distribution of the master equation 
is independent of the additional spin $\varepsilon$:
	\begin{equation}
	\forall(\bmath{\sigma},\varepsilon),\quad
	p(\bmath{\sigma},\varepsilon,t)\to\frac12\pi(\bmath{\sigma})
	\quad{\rm as}\quad t\to\infty.
	\label{eq:converge}
	\end{equation}
Therefore, the balance condition (BC) is given by  
	\begin{eqnarray}
	&&
	\sum_j\Big[
	w_j(\bmath{\sigma},\varepsilon)\pi(\bmath{\sigma})
	-w_j(F_j\bmath{\sigma},\varepsilon)\pi(F_j\bmath{\sigma})
	\Big]\qquad
	\notag \\ && \qquad 
	+\big[
	\lambda(\bmath{\sigma},\varepsilon)
	-\lambda(\bmath{\sigma},-\varepsilon)
	\big]\pi(\bmath{\sigma})
	=0. 
	\label{eq:SBC}
	\end{eqnarray} 
In order to determine the transition probabilities
$w_j(\bmath{\sigma},\varepsilon)$ and $\lambda(\bmath{\sigma},\varepsilon)$ 
satisfying BC in Eq.~(\ref{eq:SBC}), 
we impose an alternative condition:
	\begin{equation}
	w_j(\bmath{\sigma},\varepsilon)\pi(\bmath{\sigma})=
	w_j(F_j\bmath{\sigma},-\varepsilon)\pi(F_j\bmath{\sigma}).
	\label{eq:SDBC}
	\end{equation}
This condition is referred as the skew detailed 
balance condition~\cite{Turitsyn} which 
requires a local balance of stochastic flows from a state 
$(\bmath{\sigma},\varepsilon)$ to $(F_j\bmath{\sigma},\varepsilon)$ 
and that from $(F_j\bmath{\sigma}, -\varepsilon)$ to
$(\bmath{\sigma},-\varepsilon)$.
Under SDBC, 
the condition for $\lambda(\bmath{\sigma},\varepsilon)$ 
is obtained from BC in Eq.~(\ref{eq:SBC}) as
	\begin{equation}
	\lambda(\bmath{\sigma},\varepsilon)-
	\lambda(\bmath{\sigma},-\varepsilon)=
	\sum_j\Big[
	w_j(\bmath{\sigma},-\varepsilon)-
	w_j(\bmath{\sigma},\varepsilon)
	\Big].
	\label{eq:lambda}
	\end{equation}  
The condition in Eq.~(\ref{eq:lambda}) combined with SDBC ensures 
that the probability distribution converges to the equilibrium 
distribution in the Markov chain.

There still remains the degree of freedom for determining 
the transition probability $w_j(\bmath{\sigma},\varepsilon)$ 
even when SDBC is imposed.
In the present work, we choose a Glauber like 
transition probability 
	\begin{equation}
	w_j(\bmath{\sigma},\varepsilon)=
	\frac12\alpha\left[1-\frac12\gamma\sigma_j
	(\sigma_{j-1}+\sigma_{j+1})\right]
	(1-\delta\varepsilon\sigma_j), 
	\label{eq:Sw_j}
	\end{equation}
where $\delta$ is a parameter 
which characterizes the deviation from DBC. 
Note that DBC is reduced to the case of $\delta=0$ 
and the range of the parameter $\delta$ is restricted to the interval 
$[-1,1]$ because of the non-negativity of 
$w_j(\bmath{\sigma},\varepsilon)$. 
The transition probability in Eq.~(\ref{eq:Sw_j}) is equivalent to 
the heat-bath transition probability in the one-dimensional Ising model 
with virtual magnetic field $\varepsilon H$ with $\delta={\rm tanh}\beta H$.
However, it should be reminded that 
SDBC leads to the equilibrium distribution of the system without 
the magnetic field given by Eq.~(\ref{eq:converge}).

Using the transition probability in Eq.~(\ref{eq:Sw_j}), 
the condition in Eq.~(\ref{eq:lambda}) is rewritten as 
	\begin{equation}
	\lambda(\bmath{\sigma},\varepsilon)-
	\lambda(\bmath{\sigma},-\varepsilon)=
	\alpha\delta(1-\gamma)N\varepsilon m(\bmath{\sigma}). 
	\label{eq:lambda2}
	\end{equation}
There are variations in $\lambda(\bmath{\sigma},\varepsilon)$ satisfying 
the condition of Eq.~(\ref{eq:lambda2}). 
In this work, we discuss the following three types of the transition 
probabilities:
	\begin{eqnarray}
	\lambda(\bmath{\sigma},\varepsilon)
	&\!\!\!\!\!=\!\!\!\!\!&
	\sum_j w_j(\bmath{\sigma},-\varepsilon)
	\notag \\ 
	&\!\!\!\!\!=\!\!\!\!&
	\frac12\alpha N\Big[
	1-\gamma u(\bmath{\sigma})+\delta(1-\gamma)\varepsilon 
	m(\bmath{\sigma})\Big],
	\label{eq:b}
	\end{eqnarray}
	\begin{equation}
	\lambda(\bmath{\sigma},\varepsilon)=
	\frac12\alpha\delta(1-\gamma)N\Big(
	1+\varepsilon m(\bmath{\sigma})\Big),
	\label{eq:a}
	\end{equation}
and
	\begin{equation}
	\lambda(\bmath{\sigma},\varepsilon)=
	\max\Big[0,~\alpha\delta(1-\gamma)N\varepsilon 
	m(\bmath{\sigma})\Big], 
	\label{eq:c}
	\end{equation}
where $u(\bmath{\sigma})=\frac1N\sum_j\sigma_j\sigma_{j+1}$. 
We refer the first transition probability of Eq.~(\ref{eq:b})
as Sakai-Hukushima 1 (SH$_1$) type. 
Because the condition of Eq.~(\ref{eq:lambda}) means that a 
difference between $\lambda(\bmath{\sigma},\varepsilon)$ and 
$\lambda(\bmath{\sigma},-\varepsilon)$ is equal to that 
of the summation of $w_j(\bmath{\sigma},\varepsilon)$, 
$\lambda(\bmath{\sigma},\varepsilon)$ of the SH$_1$ type is 
assigned to one of the summation terms. 
The second transition probability of Eq.~(\ref{eq:a}), 
referred as Sakai-Hukushima 2 (SH$_2$) type, is obtained by 
imposing that $\lambda(\bmath{\sigma},\varepsilon)$ is 
a linear function of $\varepsilon$. 
Then, the non-negativity of $\lambda(\bmath{\sigma},\varepsilon)$ further 
restricts the range of $\delta$ to $[0,1]$. 
This transition probability is specific to the Ising model in one dimension.
The transition probability of Eq.~(\ref{eq:c}) is constructed by 
allocating all of the right hand side of Eq.~(\ref{eq:lambda}) 
to either $\lambda(\bmath{\sigma},\varepsilon)$ or 
$\lambda(\bmath{\sigma},-\varepsilon)$.
Since this transition probability has been proposed originally by 
Turitsyn et al.~\cite{Turitsyn}, we call it 
Turitsyn-Chertkov-Vucelja (TCV) type.

\subsection{Exact solution in the SH$_1$ type}
\label{3.2}

We discuss time evolution of the magnetization density under 
the SH$_1$ type transition probability 
in this subsection.
An expectation of an observable $A=A(\bmath{\sigma},\varepsilon)$ 
of this system at time $t$ is redefined from Eq.~(\ref{eq:expectation_t}) to
	\begin{equation}
	\left<A(t)\right>:=
	\sum_{\varepsilon=\pm 1}\sum_{\bmath{\sigma}}
	A(\bmath{\sigma},\varepsilon)p(\bmath{\sigma},\varepsilon,t). 
	\end{equation}
From the master equation in Eq.~(\ref{eq:Smaster}),  
differential equations for the magnetization density 
and the expectation of the additive spin $\varepsilon$ are obtained as 
	\begin{equation}
	\frac1\alpha\frac{d}{dt}\left<m(t)\right>=
	-(1-\gamma)\left<m(t)\right>+\delta\left<\varepsilon(t)\right>
	-\delta\gamma\left<\varepsilon u(t)\right>,
	\label{eq:Sm}
	\end{equation}
	\begin{equation}
	\frac1\alpha\frac{d}{dt}\left<\varepsilon(t)\right>=
	-\delta(1-\gamma)N\left<m(t)\right>-N\left<\varepsilon(t)\right>
	+\gamma N\left<\varepsilon u(t)\right>, 
	\label{eq:Su}
	\end{equation}
respectively.
By combining Eqs.~(\ref{eq:Sm}) and (\ref{eq:Su}), we have
	\begin{equation}
	\frac1\alpha\frac{d}{dt}\left<m(t)\right>=
	-(1+\delta^2)(1-\gamma)\left<m(t)\right>, 
	\label{eq:SM}
	\end{equation}
for large $N$. Thus, we obtain
	\begin{equation}
	\left<m(t)\right>=\left<m(0)\right>
	\exp[-\alpha(1+\delta^2)(1-\gamma)t], 
	\label{eq:SM2}
	\end{equation}
and the relaxation time is given as
	\begin{equation}
	\tau=\frac1{\alpha(1+\delta^2)(1-\gamma)}. 
	\label{eq:Srelax}
	\end{equation}

The autocorrelation function is calculated as well.
Let $p(\bmath{\sigma},\varepsilon,t+t_{\omega}|
\bmath{\sigma}',\varepsilon',t_{\omega})$
be a conditional probability that we find a state 
$(\bmath{\sigma},\varepsilon)$ at elapsed time $t$ after an equilibrium state 
$(\bmath{\sigma}',\varepsilon')$ is given at $t_{\omega}$. 
Then, Eq.~(\ref{eq:A_eqA(t)}) is extended to 
	\begin{eqnarray}
	\left<A_{\rm eq}A(t)\right>
	&\!\!\!=\!\!\!&
	\sum_{\varepsilon,\varepsilon'}
	\sum_{\bmath{\sigma},\bmath{\sigma}'}
	A(\bmath{\sigma}',\varepsilon')
	\frac12\pi(\bmath{\sigma}')
	A(\bmath{\sigma},\varepsilon)
	\notag \\ && \qquad \times
	p(\bmath{\sigma},\varepsilon,t+t_{\omega}|
	\bmath{\sigma}',\varepsilon',t_{\omega}).
	\end{eqnarray}
Therefore, we have 
	\begin{equation}
	C_{\rm eq}(t;m)=\exp[-\alpha(1+\delta^2)(1-\gamma)t], 
	\label{eq:SMC}
	\end{equation}
and the integrated autocorrelation time defined by 
Eq.~(\ref{eq:autocorrelation_time}) is obtained
	\begin{equation}
	\tau_{{\rm int},m}=\frac1{\alpha(1+\delta^2)(1-\gamma)}. 
	\label{eq:SMtauint}
	\end{equation}
As in the case of DBC discussed in the previous section, 
the relaxation time and the integrated autocorrelation time of the 
magnetization density coincide with each other.
It is shown analytically that they are reduced by introducing SDBC 
in the case of the SH$_1$ type.  
Interestingly, it is found that the DBC point $(\delta=0)$ 
is the worst efficient in this case. 
The reduction of the relaxation time from DBC is constant and 
independent of temperature. Therefore, the dynamical exponent does not change 
from that in DBC by using this type of transition probability satisfying SDBC.

\subsection{Approximate analyses in the SH$_2$ type}

In this subsection, we discuss another transition probability 
$\lambda(\bmath{\sigma},\varepsilon)$ given by Eq.~(\ref{eq:a}).
The differential equation for the expectation of the additive spin $\varepsilon$ 
is replaced from Eq.~(\ref{eq:Su}) to
	\begin{equation}
	\frac1\alpha\frac{d}{dt}\left<\varepsilon(t)\right>=
	-\delta(1-\gamma)N\left<m(t)\right>
	-\delta(1-\gamma)N\left<\varepsilon(t)\right>,
	\label{eq:Su2}
	\end{equation}
while that for the magnetization density is identical to Eq.~(\ref{eq:Sm}). 
It is difficult to solve the differential equation
because of the existence of the higher order term $\left<\varepsilon u(t)\right>$ 
in Eq.~(\ref{eq:Sm}), which differs from the case of SH$_1$ type.  
It is reasonably assumed 
	\begin{equation}
	\left<\varepsilon u(t)\right>\simeq
	\left<\varepsilon(t)\right>\left<u\right>_{\rm eq},
	\label{eq:assumption}
	\end{equation}
when the system is in the vicinity of the equilibrium state. 
Under the assumption, Eq.~(\ref{eq:Sm}) is rewritten as
	\begin{equation}
	\frac1{\alpha}\frac{d}{dt}\left<m(t)\right>=
	-(1-\gamma)\left<m(t)\right>
	+\delta\sqrt{1-\gamma^2}\left<\varepsilon(t)\right>,
	\label{eq:Sm2}
	\end{equation}
where $\left<u\right>_{\rm eq}={\rm tanh}\beta J=(1-\sqrt{1-\gamma^2})/\gamma$
for large $N$. By combining Eqs.~(\ref{eq:Su2}) and (\ref{eq:Sm2}), 
we obtain
	\begin{equation}
	\left<m(t)\right>=\left<m(0)\right>
	\exp[-\alpha(1-\gamma+\delta\sqrt{1-\gamma^2})t].
	\label{eq:SM3}
	\end{equation}
Note that this result could be valid 
if the initial state is near the equilibrium point. 
Using Eq.~(\ref{eq:SM3}), we have the autocorrelation function as 
	\begin{equation}
	C_{\rm eq}(t;m)=
	\exp[-\alpha(1-\gamma+\delta\sqrt{1-\gamma^2})t],
	\label{eq:SM4}
	\end{equation}
and the integrated autocorrelation time is
	\begin{equation}
	\tau_{{\rm int},m}=\frac1{\alpha(1-\gamma+\delta\sqrt{1-\gamma^2})}.
	\label{eq:Stauint}
	\end{equation}
This expression is different from that obtained in the SH$_1$ type. 
The relaxation time is not an even function of the parameter $\delta$. 
However the parameter takes the value of $[0,1]$ in this type. 
Hence, it is also found that the case of DBC with $\delta=0$ gives 
the worst efficiency in the parameter range. 
Further, the gain from the DBC point significantly depends on temperature 
and the dynamical critical exponent $z$ is down to $1$ in this type of transition 
probability of $\varepsilon$ flip. While some numerical works suggest 
the reduction of the dynamical critical exponent by 
using non reversible transition 
probability~\cite{Fernandes, Turitsyn}, 
the present study shows analytically that the specific type of 
irreversible transition probability makes the relaxation of 
the magnetization density accelerated and 
changes the dynamical critical phenomena.

\subsection{Linear analyses of time evolution near equilibrium point}
\label{3.4}

As seen in the previous section, an irreversible transition probability 
yields the reduction of dynamical critical exponent. Non-local update such as 
cluster algorithm~\cite{Swendsen, Wolff} often leads to such acceleration 
of dynamics.
It should be noted that even a local spin update changes the dynamical properties 
of system by using the irreversible transition probability. 
However, the irreversible transition probability does not always make 
the significant changes as seen in \S\ref{3.2}.
It turns out that the transition probability for $\varepsilon$ flip 
plays an important role for the dynamics of the spin configuration.

In this subsection, we study dynamical trajectory of the magnetization density 
and the expectation of the additive Ising spin near equilibrium point 
in the case of the SH$_1$ type and SH$_2$ type.  
Under the assumption in Eq.~(\ref{eq:assumption}), the differential equations 
of these estimators can be solved for the SH$_1$ type 
in a thermodynamical limit
	\begin{figure}[t]
		\begin{center}
		\includegraphics[width=.5\columnwidth,clip]{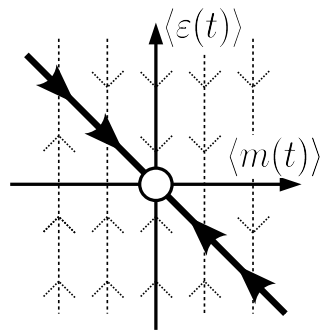}
		\label{fig:behavior_of_estimators}
		\caption{Schematic picture of dynamical behavior of 
		$\left<m(t)\right>$ and $\left<\varepsilon(t)\right>$ 
		in the vicinity of the equilibrium point at origin 
		for the transition probability of the SH$_1$ and SH$_2$ types. 
		Bold solid line represents a slowest mode of the dynamics 
		along which the set of estimators approaches to the equilibrium point 
		soon after starting from an initial point 
		$(\left<m(0)\right>,\left<\varepsilon(0)\right>)$.}		
		\end{center}
	\end{figure}
 	\begin{equation}
	\left\{
		\begin{array}{l}
		\left<m(t)\right>=\left<m(0)\right>\exp[-\alpha(1+\delta^2)(1-\gamma)t],
		\\
		\displaystyle
		\left<\varepsilon(t)\right>=-\delta\sqrt{\frac{1-\gamma}{1+\gamma}}
		\left<m(0)\right>\exp[-\alpha(1+\delta^2)(1-\gamma)t],
		\end{array}
	\right.
	\label{eq:SH_1}
	\end{equation}
and for the SH$_2$ type
	\begin{equation}
	\left\{
		\begin{array}{l}
		\left<m(t)\right>=\left<m(0)\right>
		\exp[-\alpha(1-\gamma+\delta\sqrt{1-\gamma^2})t],
		\\
		\displaystyle
		\left<\varepsilon(t)\right>=-\left<m(0)\right>
		\exp[-\alpha(1-\gamma+\delta\sqrt{1-\gamma^2})t].
		\end{array}
	\right.
	\label{eq:SH_2}
	\end{equation}
These solutions are represented as a dynamical trajectory in the parameter 
space of $\left<m(t)\right>$ and $\left<\varepsilon(t)\right>$, 
where the equilibrium point is the origin $(0,0)$.
Using linear analysis at the equilibrium point, it is found 
that these expectations converge to the equilibrium point $(0,0)$ 
along by a straight line, which can be regarded as an eigenvector of the 
slowest mode of the dynamics.
Figure 1 
shows a schematic picture of the trajectory. 
In the case of DBC, $\left<\varepsilon(t)\right>$ is an irrelevant parameter 
and thus the magnetization density $\left<m(t)\right>$ 
converges to zero along by the horizontal axis.
A finite slope of the asymptotic line is a consequence of the irreversible 
transition probability. 
In fact, the solutions in Eqs.~(\ref{eq:SH_1}) and (\ref{eq:SH_2}) provide 
the estimate of the slope as $-\delta\sqrt{(1-\gamma)/(1+\gamma)}$ for 
the SH$_1$ type and $-1$ for the SH$_2$ type. 
The slope for the SH$_1$ type decreases with temperature decreasing 
and eventually goes to zero at zero temperature, $\gamma=1$.
Namely, the dynamics near zero temperature is essentially 
equivalent to that in DBC.
Presumably, this is the reason why the transition probability of the SH$_1$ 
type does not change the dynamical critical phenomena.
On the other hand, the slope for the SH$_2$ type is independent of 
temperature and quite different from that for DBC. 
This implies that the relaxation dynamics to the equilibrium state with SDBC 
is accelerated by using the extended state including the additional 
$\varepsilon$ spin.

\section{Monte Carlo simulations}
\label{simulations}

\begin{figure*}[t]
	\begin{center}
		\begin{tabular}{ccc}
			\begin{minipage}{.3\textwidth}
			\includegraphics[width=.95\columnwidth,clip]{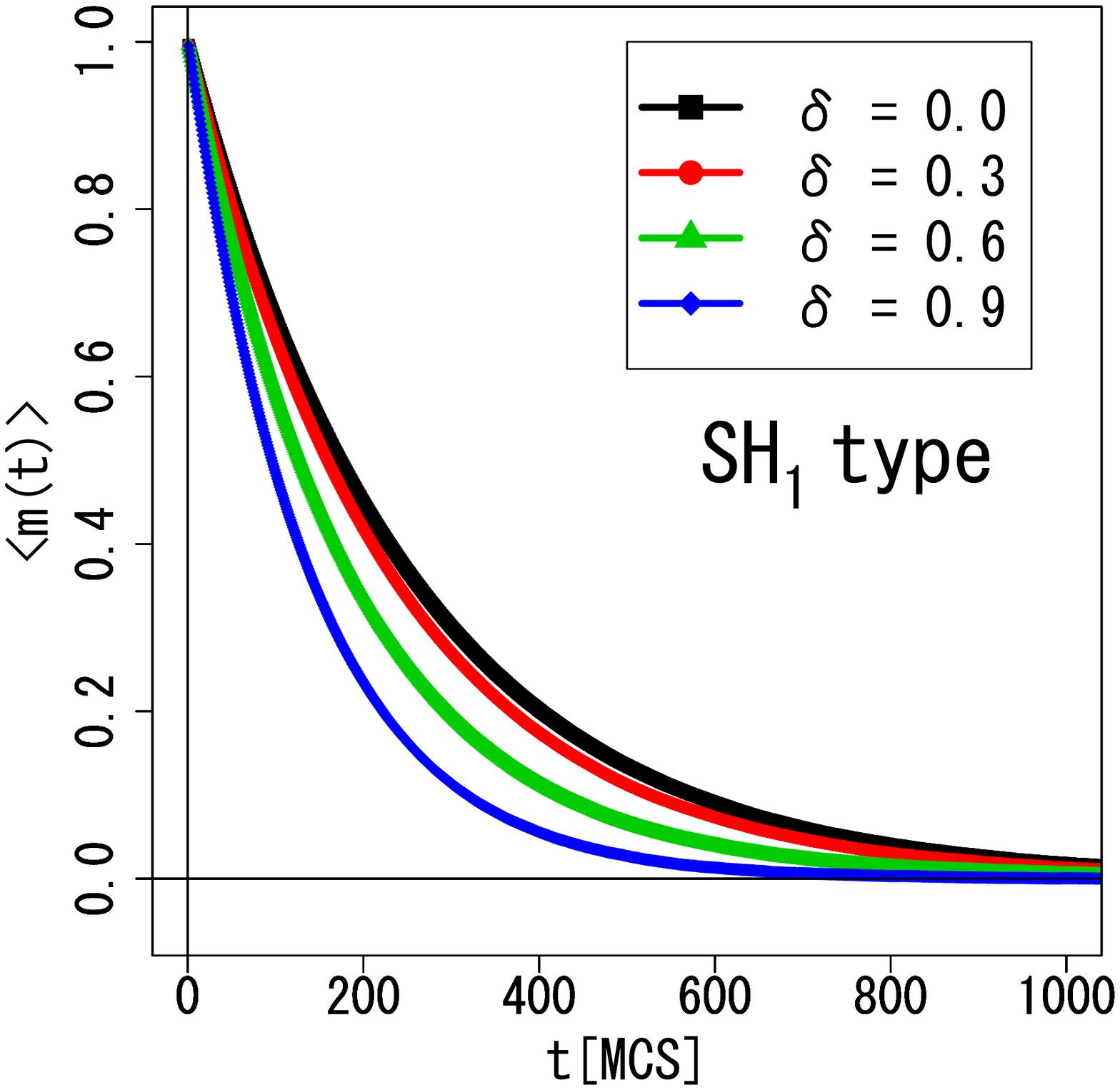}	
			\end{minipage}
			&
			\begin{minipage}{.3\textwidth}
			\includegraphics[width=.95\columnwidth,clip]{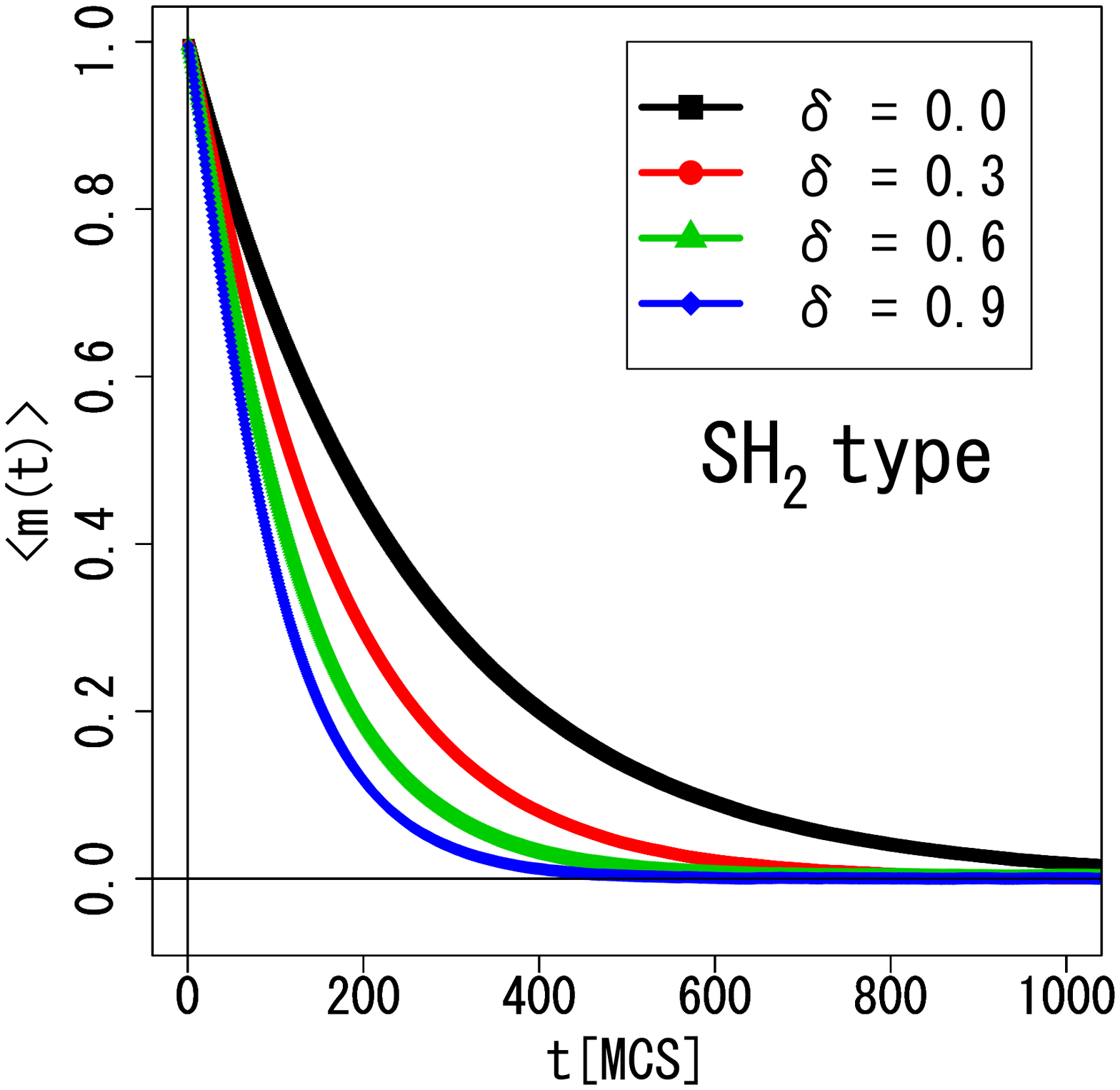}	
			\end{minipage}			
			&
			\begin{minipage}{.3\textwidth}
			\includegraphics[width=.95\columnwidth,clip]{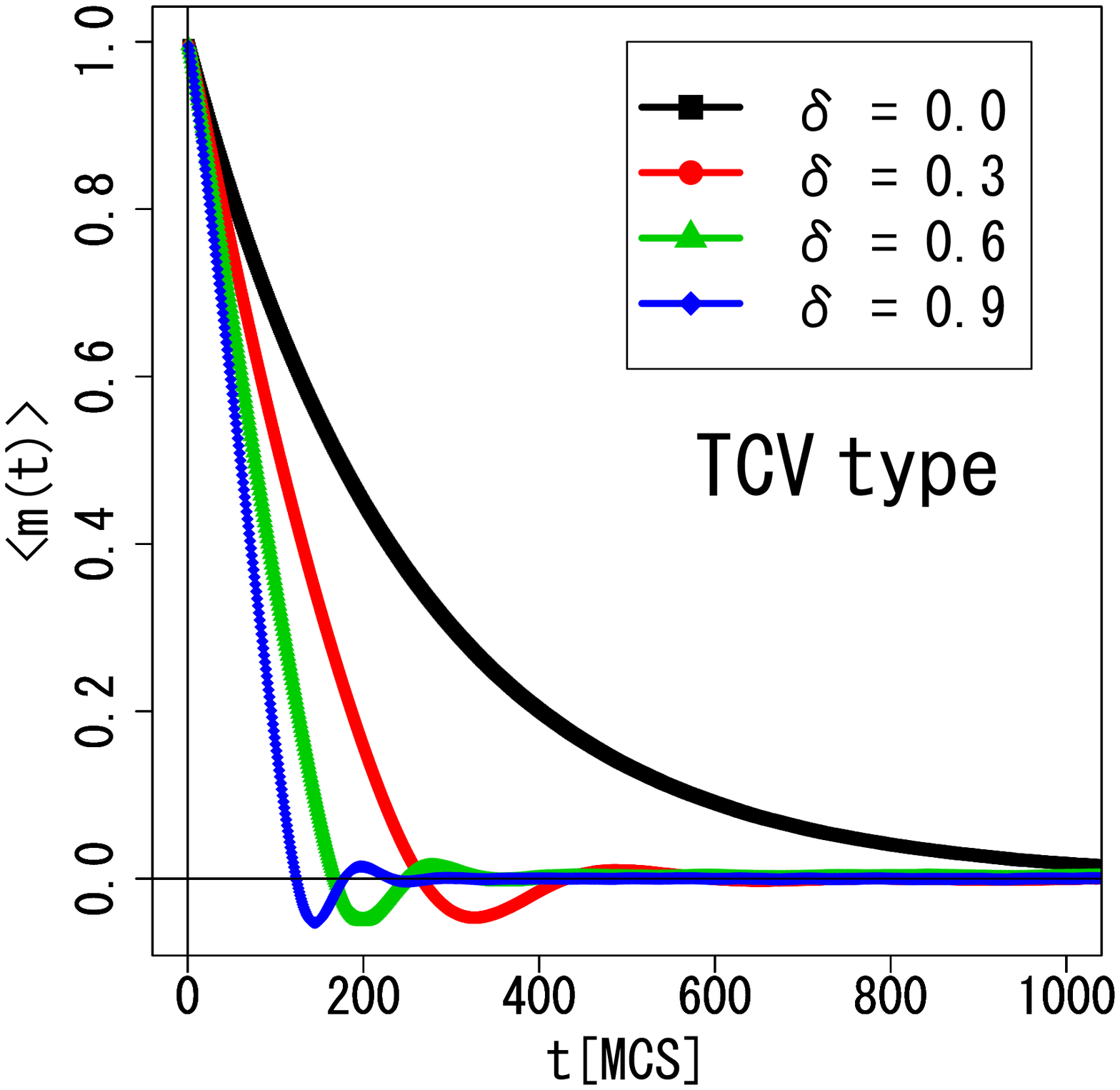}	
			\end{minipage}			
		\end{tabular}
	\end{center}
\caption{(Color online) Time evolution of 
the magnetization density in the one-dimensional Ising model 
for different values of $\delta$. 
The chosen values of parameter in the simulations are
$N=2^7$, $\alpha=10^{-2}$, $\gamma=0.6$, $N_{\rm ens}=10^5$.
The transition probability used is the SH$_1$, SH$_2$, and TCV types 
from left to right, respectively.}
\label{fig:irN128_r6_m}
	\begin{center}
		\begin{tabular}{ccc}
			\begin{minipage}{.3\textwidth}
			\includegraphics[width=.95\columnwidth,clip]{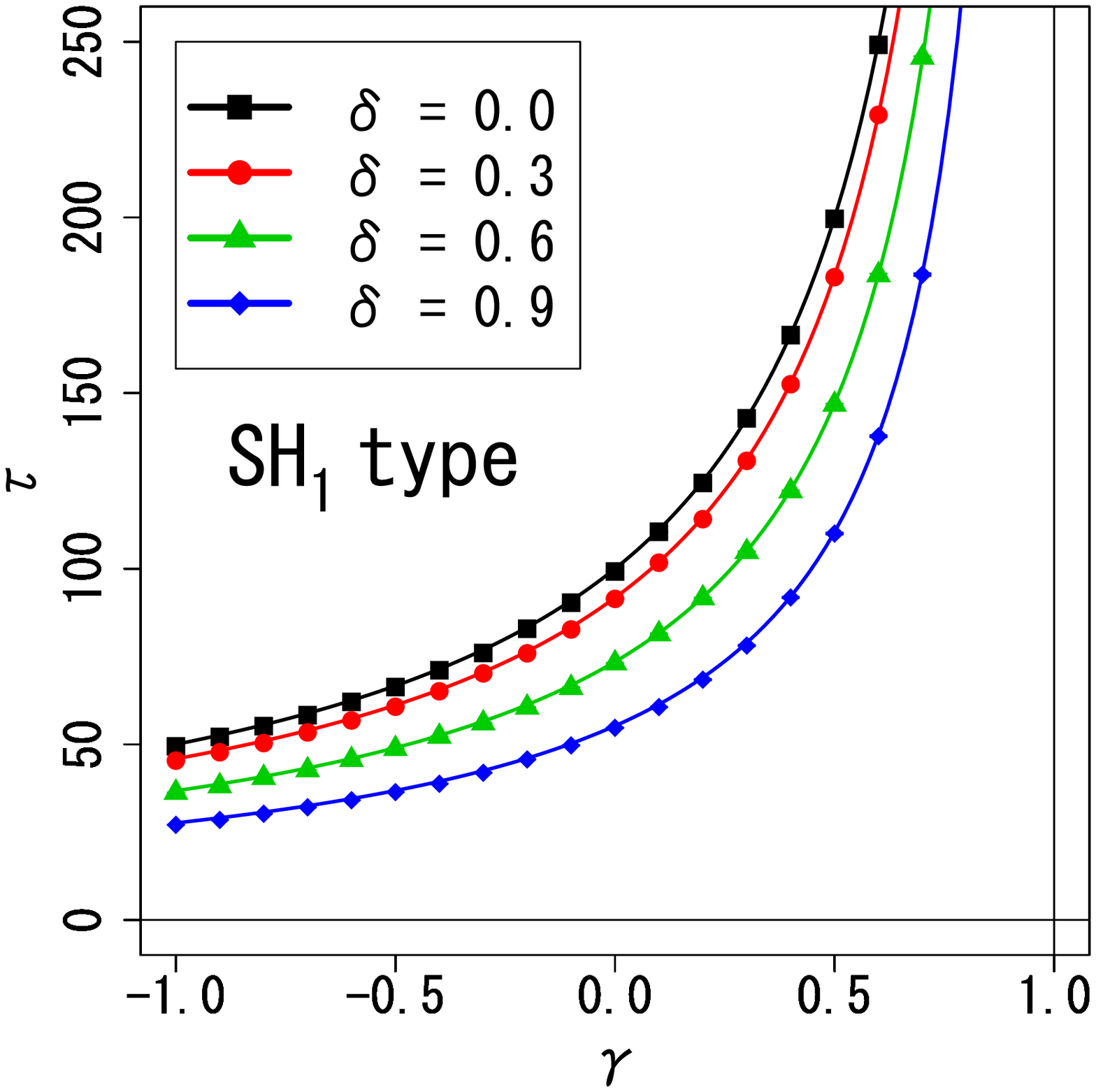}	
			\end{minipage}
			&
			\begin{minipage}{.3\textwidth}
			\includegraphics[width=.95\columnwidth,clip]{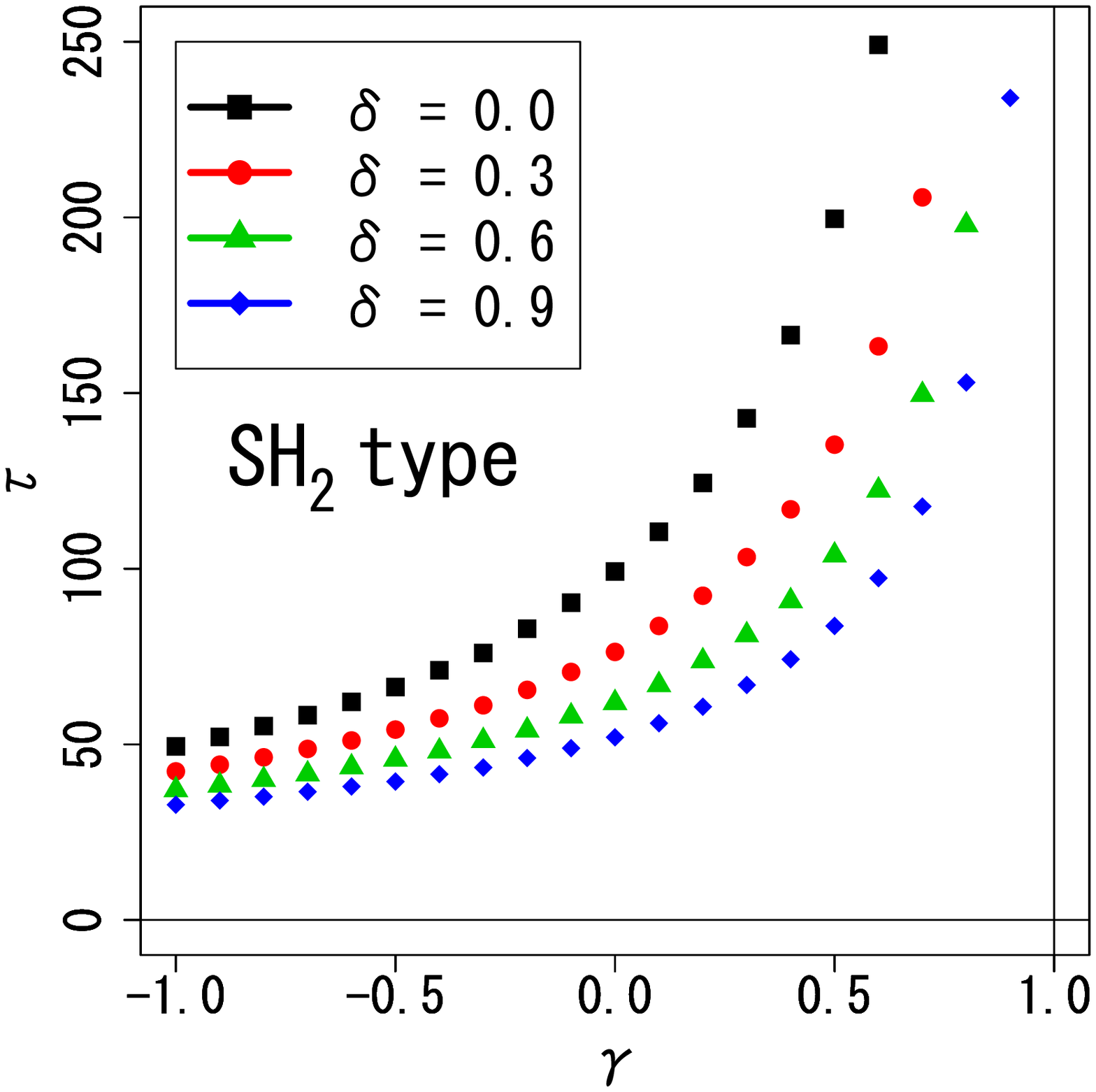}	
			\end{minipage}			
			&
			\begin{minipage}{.3\textwidth}
			\includegraphics[width=.95\columnwidth,clip]{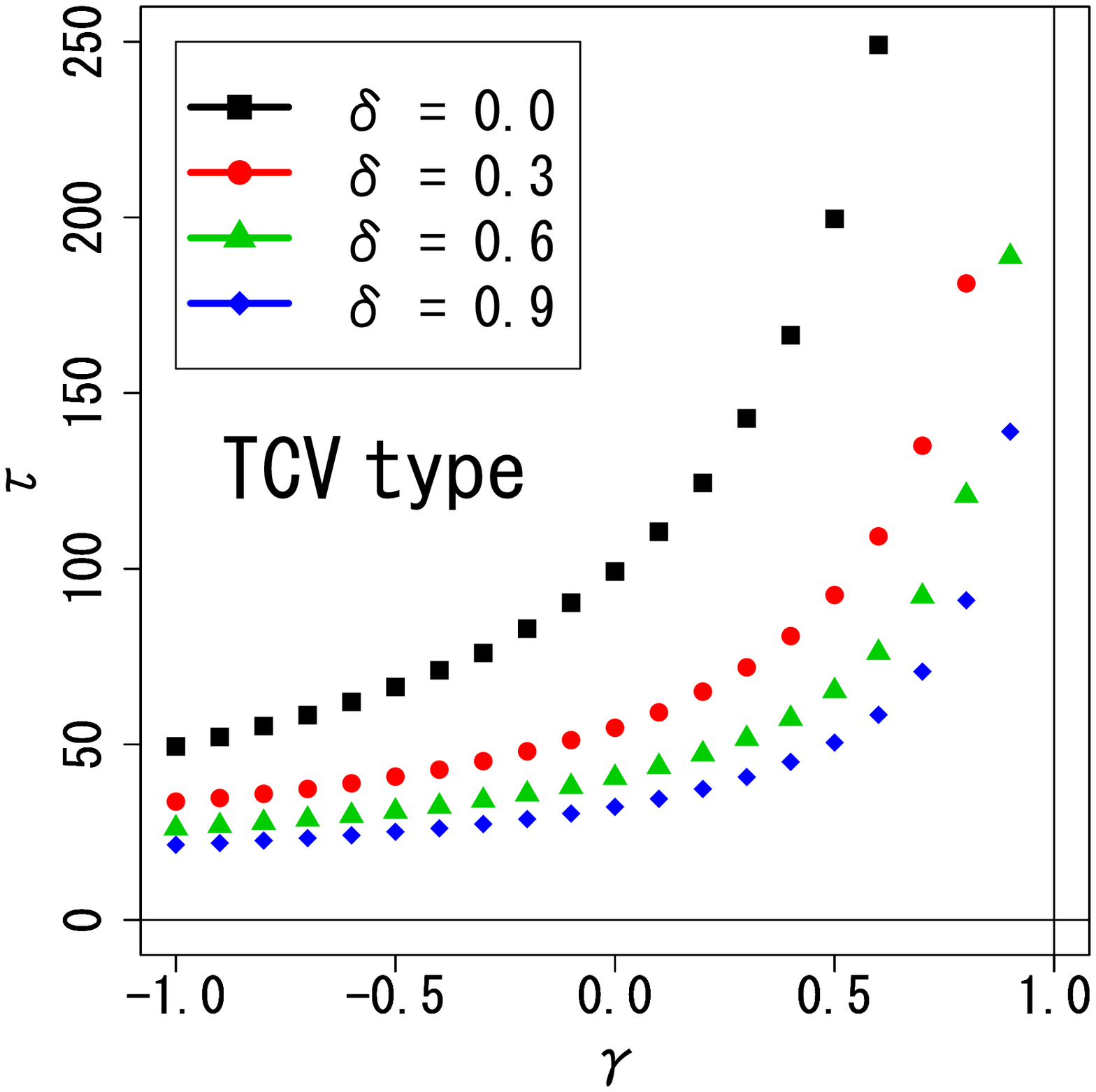}	
			\end{minipage}			
		\end{tabular}
\caption{(Color online) $\gamma$ dependence 
of the relaxation time of the magnetization density with the parameter 
$\delta$ varying. The chosen values of parameter in the simulations are 
$N=2^7$, $\alpha=10^{-2}$, $\gamma=0.6$, $N_{\rm ens}=10^5$ 
and $M=2.5\times10^3$. 
The transition probability used is the SH$_1$, SH$_2$, and TCV types 
from left to right, respectively. 
The solid lines in the left panel for the SH$_1$ type 
represent the theoretical results.}
\label{fig:tau}
 	\end{center}
\end{figure*}

In this section, we explain a procedure of the MCMC method with SDBC 
for the one-dimensional Ising model described in the previous section. 
Let $X(n)$ be a state of the system after $n$ steps.
Then, the elementary procedure of discrete time evolution 
in our simulation is as follows:
	\begin{enumerate}[(a)]
	\item
	Set an initial condition $X(0)$ arbitrary.
	\item
	Suppose that the state $X(n)=(\bmath{\sigma},\varepsilon)$ 
	at time $n$ and choose a spin $\sigma_j$ from $\bmath{\sigma}$ at random.
	\item
	Accept the new state as $X(n+1)=(F_j\bmath{\sigma},\varepsilon)$ 
	with the probability $w_j(\bmath{\sigma},\varepsilon)$.
	If it is rejected, accept $X(n+1)=(\bmath{\sigma},-\varepsilon)$ 
	with the probability 
		\begin{equation}
		\Lambda(\bmath{\sigma},\varepsilon)=
		\frac{\frac1N\lambda(\bmath{\sigma},\varepsilon)}
		{1-\frac1N\sum_j w_j(\bmath{\sigma},\varepsilon)}. 
		\end{equation}
	If also rejected, set $X(n+1)=X(n)$.	
	Then, return to (b) and repeat the steps (b)--(c).
	\end{enumerate}
It is proven that these steps satisfy BC.
We consider $N$ steps of (b)--(c) as one Monte Carlo step (MCS).
In this work, the initial condition $X(0)$ is fixed as 
$\sigma_j=+1$ for all $j$ and $\varepsilon=\pm 1$ is chosen at random.

In this method, 
an expectation of an observable $A=A(\bmath{\sigma},\varepsilon)$ at time $t$
is estimated as  
	\begin{equation}
	\left<A(t)\right>\simeq\frac1{N_{\rm ens}}
	\sum_{i=1}^{N_{\rm ens}}A(\bmath{\sigma}^i(t),\varepsilon^i(t)),
	\label{eq:estimation_m}
	\end{equation}
where $(\bmath{\sigma}^i(t), \varepsilon^i(t))$ 
is a state of $i$-th trajectory at $t$-th MCS starting from 
the initial condition and 
$N_{\rm ens}$ denotes the number of simulated trajectories. 
Figure~\ref{fig:irN128_r6_m} shows the time dependence of the 
magnetization density estimated by Eq.~(\ref{eq:estimation_m}) 
for the transition probabilities discussed in the previous sections. 
The relaxation time of the magnetization density under the transition 
probability is estimated as 
	\begin{equation}
	\tau\simeq
	\frac1{M}\sum_{t=1}^{M}
	\left<m(t)\right>,
	\label{eq:estimation_tau}
	\end{equation}
where $M$ is the total number of MCS.
Figure~\ref{fig:tau} shows the $\gamma$ dependence of 
the relaxation time of the magnetization density with the parameter $\delta$ 
varying. 
From Fig.~\ref{fig:irN128_r6_m}, it turns out that the magnetization 
density decays exponentially in time and it converges to zero rapidly 
with increasing $\delta$ in the case of all three types. In particular, 
in the case of SH$_1$ type, this is completely consistent with the 
results of Eq.~(\ref{eq:SM2}) for $N$ and $\alpha$ used in the simulations. 
Moreover, the numerical estimation of relaxation time, shown in 
the left panel of Fig.~\ref{fig:tau}, 
is consistent with the theoretical estimate in Eq.~(\ref{eq:Srelax}). 
\begin{figure*}[ht]
	\begin{center}
		\begin{tabular}{ccc}
			\begin{minipage}{.3\textwidth}
			\includegraphics[width=.95\columnwidth,clip]{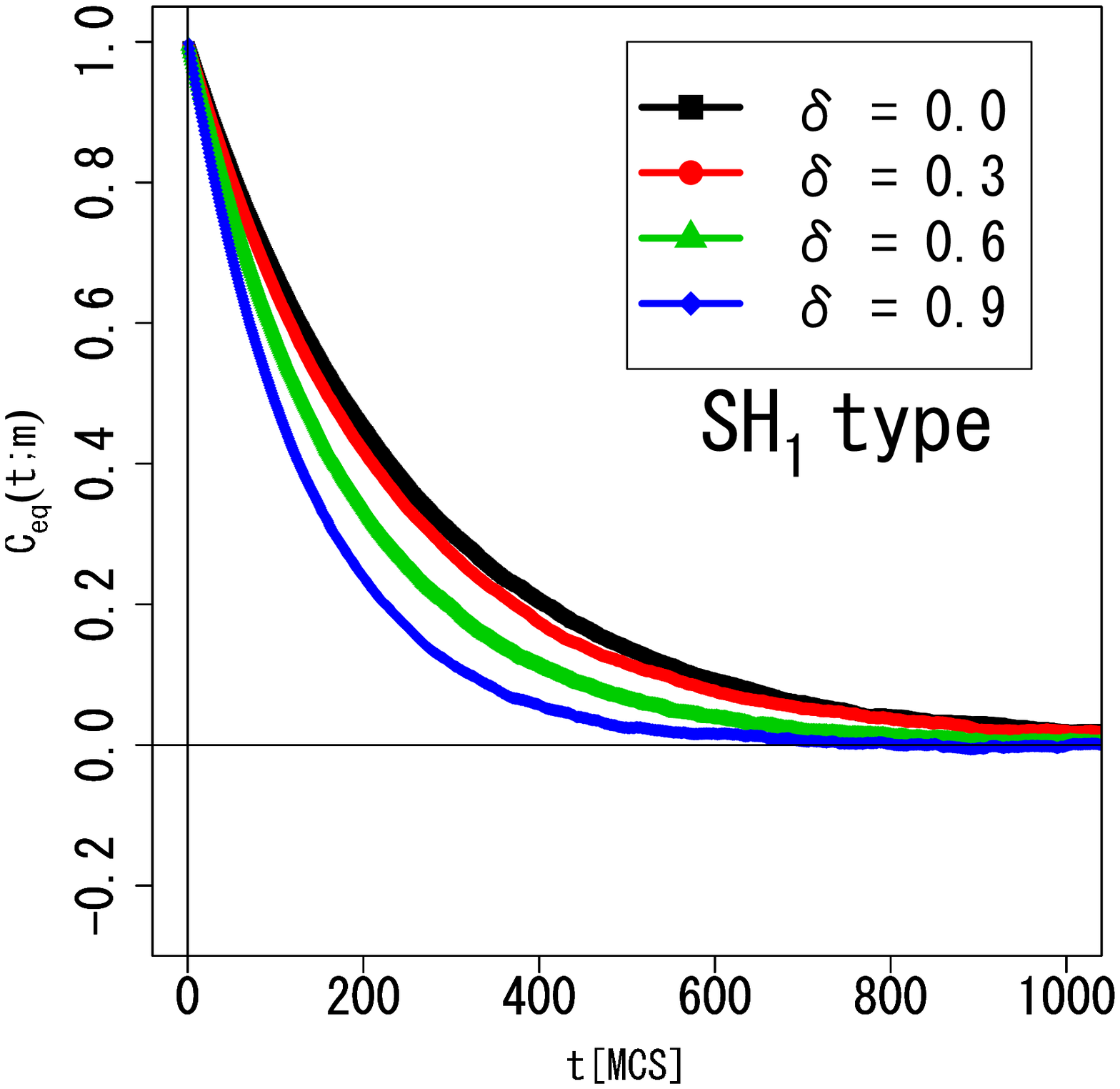}	
			\end{minipage}
			&
			\begin{minipage}{.3\textwidth}
			\includegraphics[width=.95\columnwidth,clip]{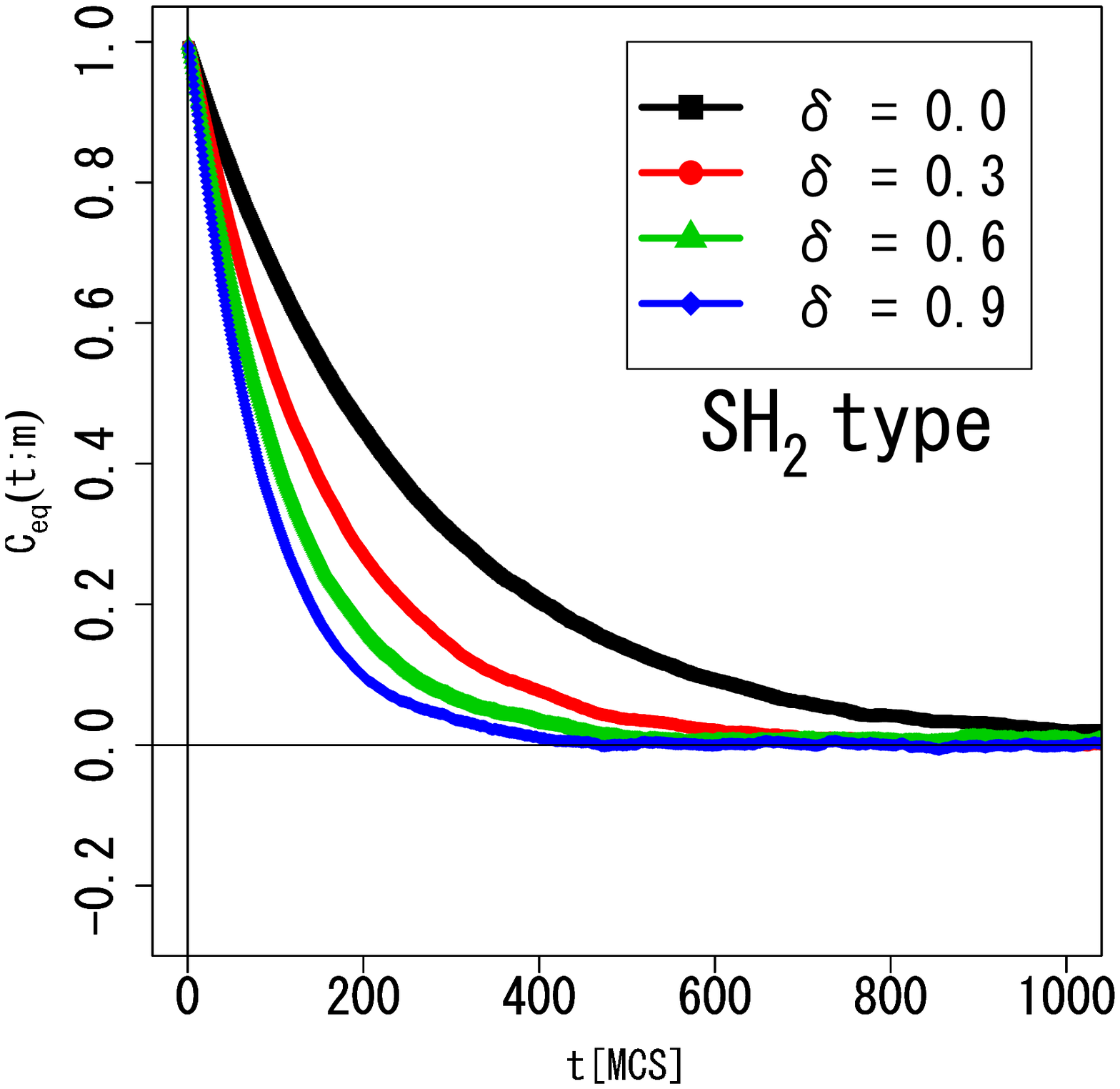}	
			\end{minipage}			
			&
			\begin{minipage}{.3\textwidth}
			\includegraphics[width=.95\columnwidth,clip]{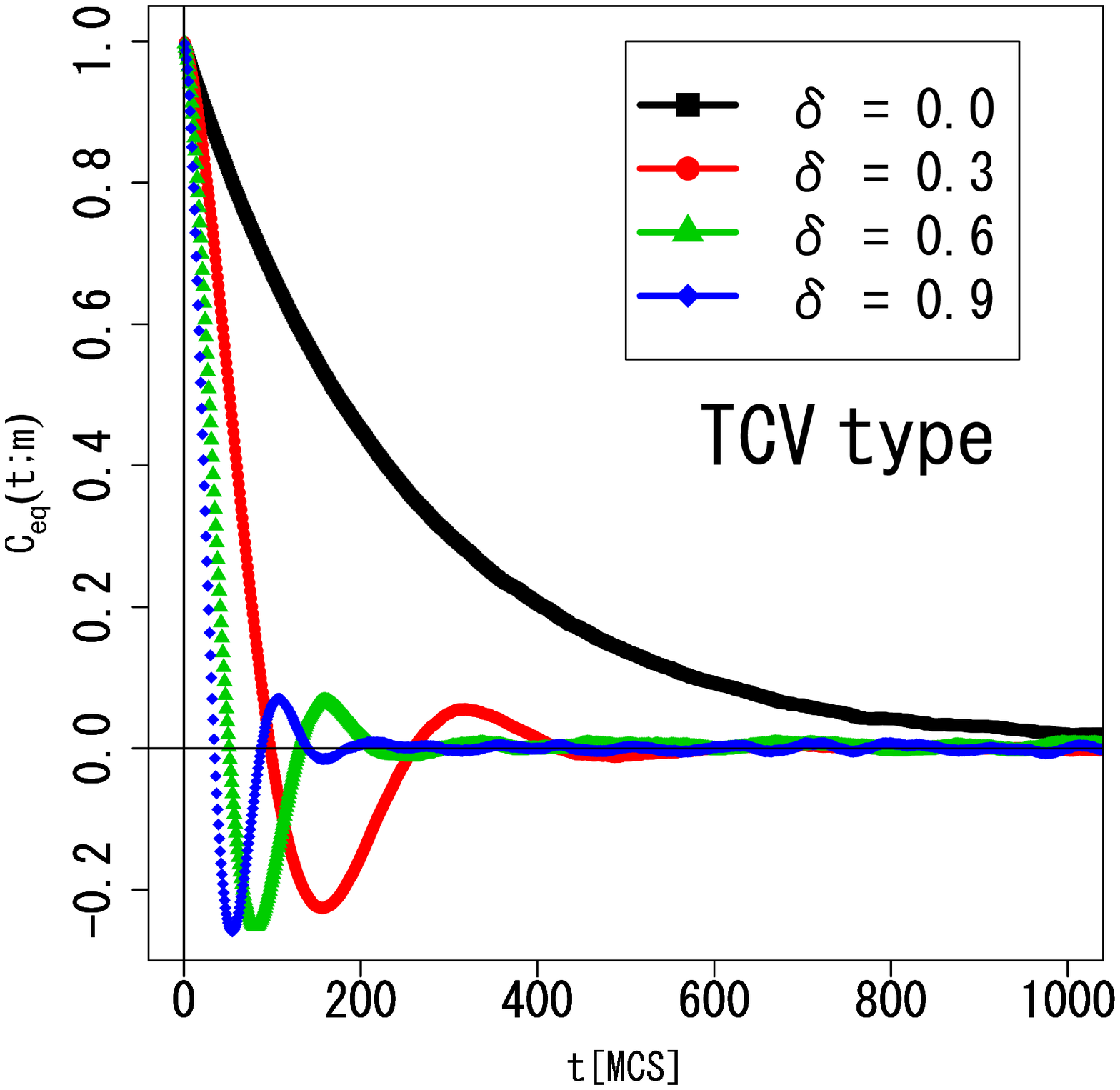}	
			\end{minipage}			
		\end{tabular}
\caption{(Color online) Time evolution of autocorrelation 
function of the magnetization density in the one-dimensional Ising model 
for different values of $\delta$. 
The chosen values of parameter in the simulations are
$N=2^7$, $\alpha=10^{-2}$, $\gamma=0.6$, $N_{\rm ens}=10^5$.
The transition probability used is the SH$_1$, SH$_2$, and TCV types 
from left to right, respectively.}
\label{fig:autocorrelation}
 	\end{center}
	\begin{center}
		\begin{tabular}{ccc}
			\begin{minipage}{.3\textwidth}
			\includegraphics[width=.95\columnwidth,clip]{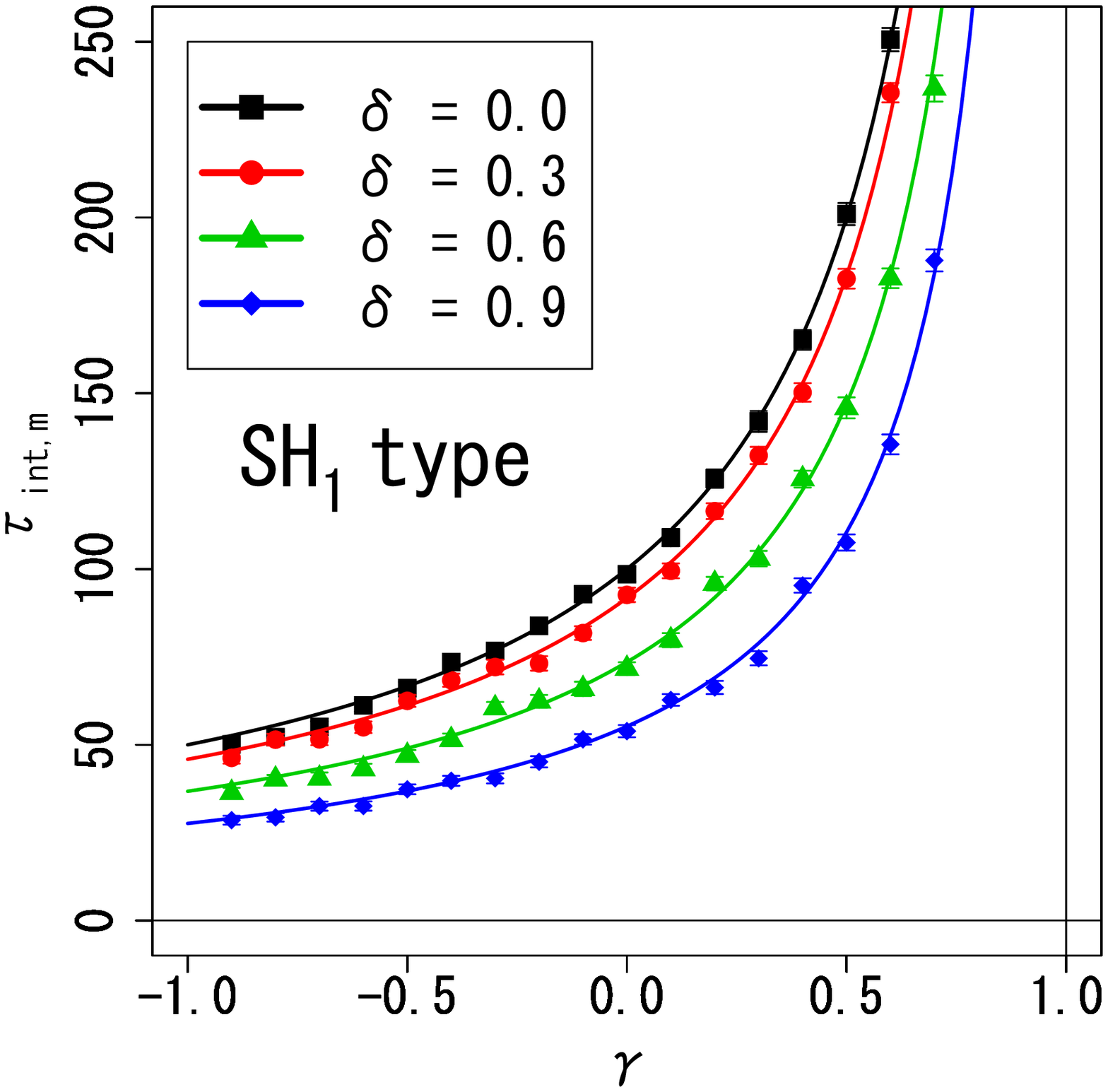}	
			\end{minipage}
			&
			\begin{minipage}{.3\textwidth}
			\includegraphics[width=.95\columnwidth,clip]{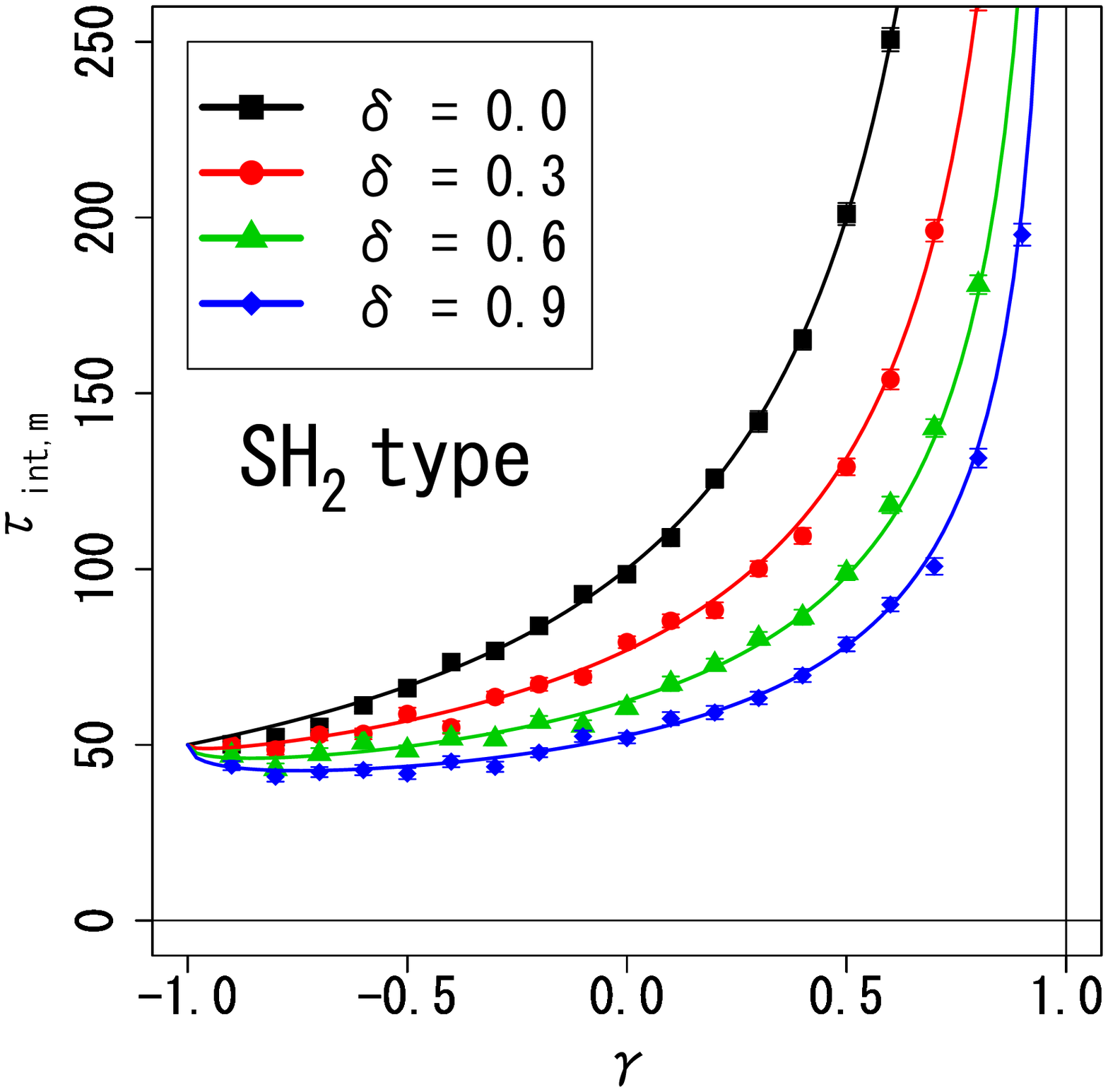}	
			\end{minipage}
			&
			\begin{minipage}{.3\textwidth}
			\includegraphics[width=.95\columnwidth,clip]{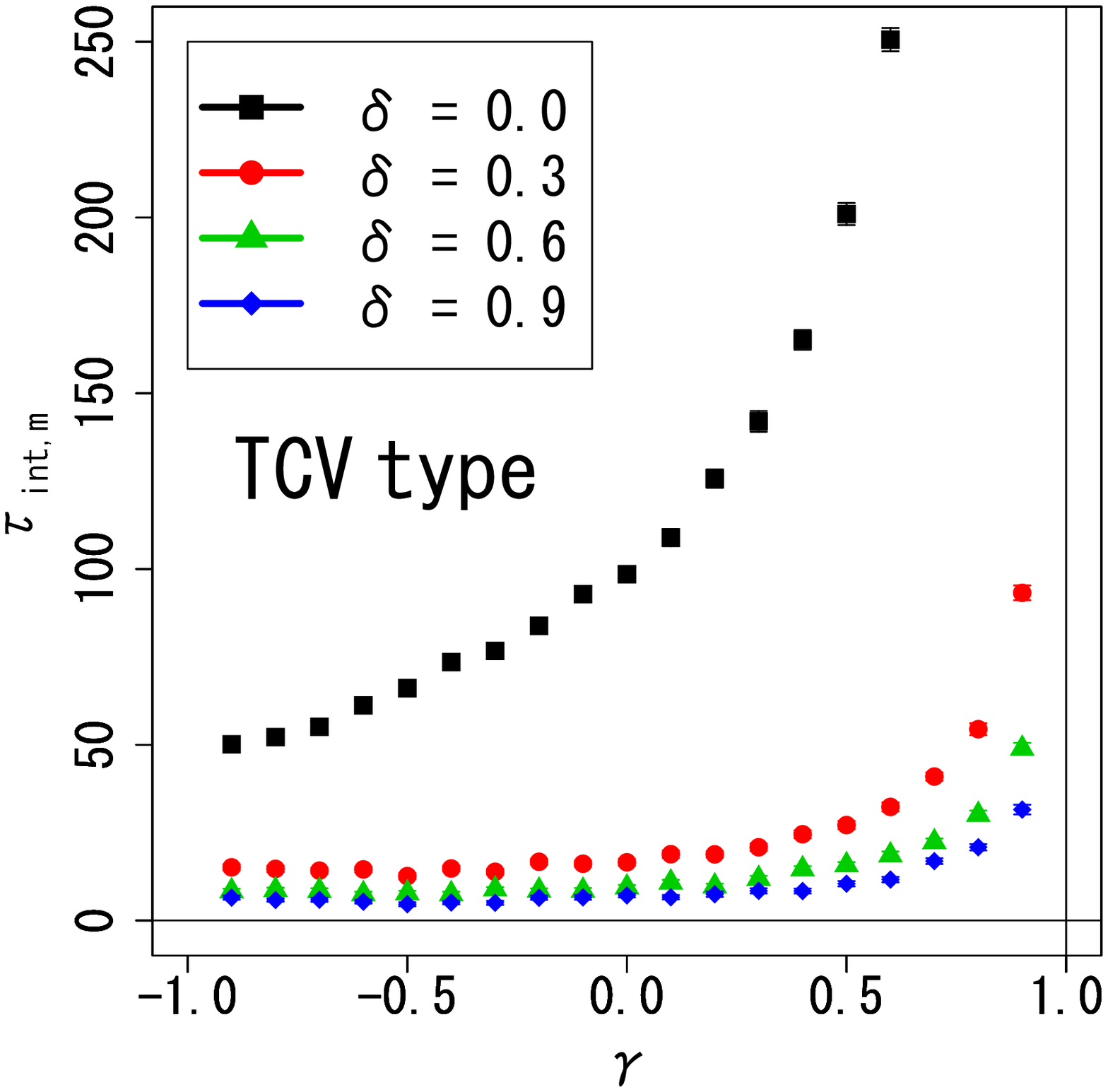}	
			\end{minipage}			
		\end{tabular}
	\caption{(Color online) $\gamma$ dependence 
	of the integrated autocorrelation time of 
	the magnetization density with the parameter 
	$\delta$ varying. The chosen values of parameter in the simulations are 
	$N=2^7$, $\alpha=10^{-2}$, $\gamma=0.6$, $N_{\rm ens}=10^5$ 
	and $M=2.5\times10^3$. 
	The transition probability used is the SH$_1$, SH$_2$, and 
	TCV types from left to right, respectively. 
	The solid lines in the left and middle panels
	represent the theoretical results.}
	\label{fig:tauint}
	\end{center}
\end{figure*}

The autocorrelation function and the integrated autocorrelation time are 
also calculated. 
Let $t_{\omega}$ be a sufficient large integer which ensures
equilibrium of the system. Then, 
the autocorrelation function of the magnetization density is estimated as 
	\begin{equation}
	\left<m_{\rm eq}m(t)\right>\simeq\frac1{N_{\rm ens}}
	\sum_{i=1}^{N_{\rm ens}}
	m(\bmath{\sigma}^i(t_{\omega}))m(\bmath{\sigma}^i(t+t_{\omega})),
	\label{eq:estimation_autocorrelation}
	\end{equation}
and the integrated autocorrelation time of the magnetization density is 
	\begin{equation}
	\tau_{{\rm int},m}\simeq
	\frac1{M}\sum_{t=1}^{M}
	\frac{\left<m_{\rm eq}m(t)\right>-\left<m\right>^2_{\rm eq}}
	{\left<m^2\right>_{\rm eq}-\left<m\right>^2_{\rm eq}},
	\label{eq:estimation_tauint}
	\end{equation}
where here $M$ is the total number of MCS after $t_{\omega}$ steps and 
the equilibrium values of $\left<m\right>_{\rm eq}$ and 
$\left<m^2\right>_{\rm eq}$ are used. Figures \ref{fig:autocorrelation} and
\ref{fig:tauint} present the numerical results estimated by 
Eqs.~(\ref{eq:estimation_autocorrelation}) and (\ref{eq:estimation_tauint}), 
respectively.
As seen in the magnetization density, the autocorrelation function also decays to 
zero exponentially in time and the relaxation is accelerated by increasing 
$\delta$ for three cases. 
MC results recover the theoretical solutions of Eqs.~(\ref{eq:SMC}) and 
(\ref{eq:SMtauint}) for the SH$_1$ type. 
Further, the approximate solution for the SH$_2$ type describes well 
$\tau_{{\rm int}, m}$ obtained by the MC simulation, including that 
$\delta$ dependence of $\tau_{{\rm int}, m}$ disappears at zero temperature 
limit with anti-ferromagnetic interaction, $\gamma=-1$. 
This confirms numerically that the transition probability of the SH$_2$ 
type changes the dynamical critical phenomena. 
For the TCV type, an oscillating behavior of the autocorrelation function is 
clearly observed in the right panel of Fig.~\ref{fig:autocorrelation}
and consequently the integrated autocorrelation time is reduced more 
significantly than that for the other types. 
This implies the existence of complex eigenmode in the relaxation dynamics, 
discussed later.

\begin{figure*}[ht]
		\begin{center}
			\begin{tabular}{ccc}
				\begin{minipage}{.3\textwidth}
				\includegraphics[width=.95\columnwidth,clip]{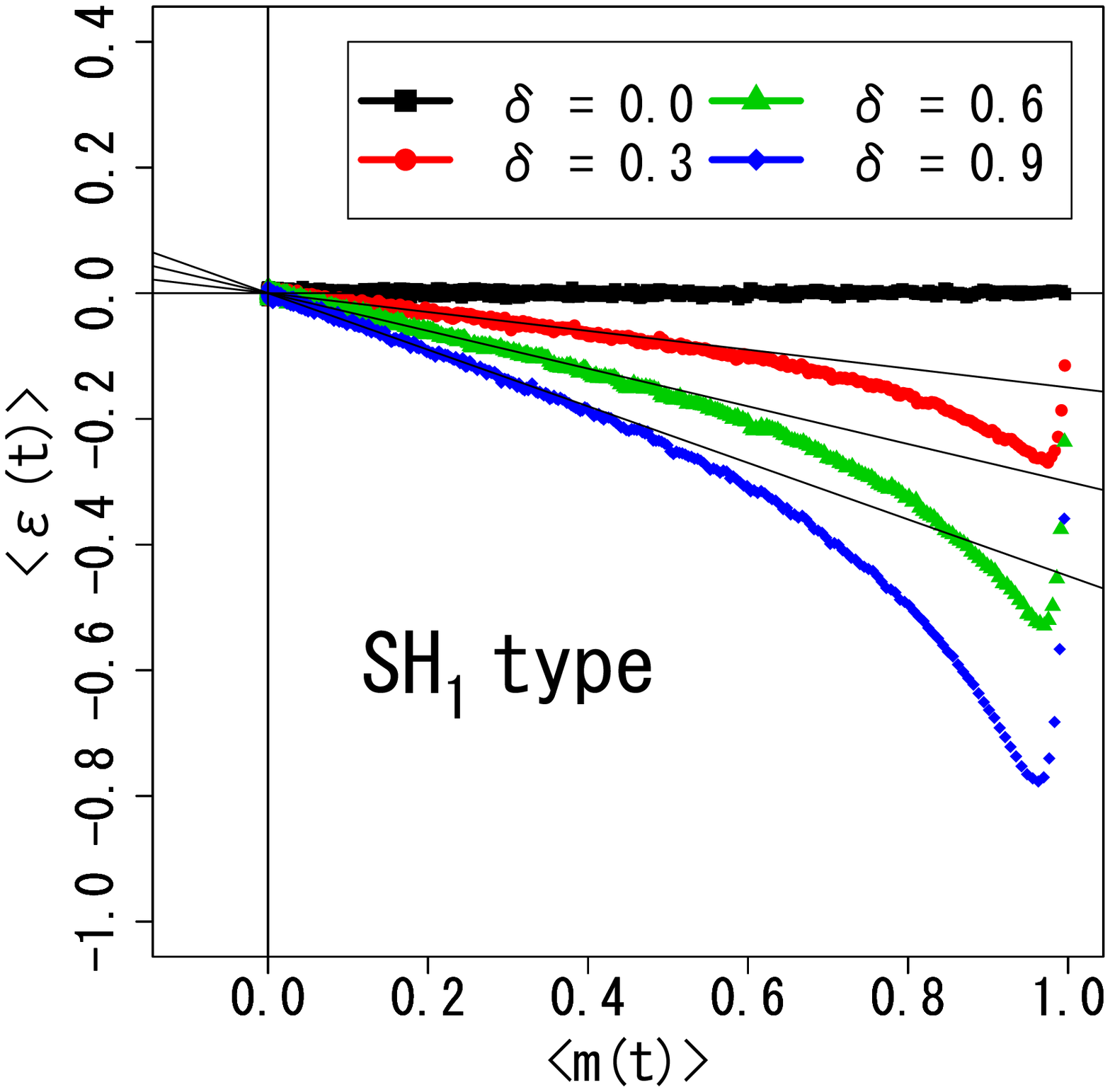}	
				\end{minipage}
				&
				\hspace{.15\textwidth}
				&
				\begin{minipage}{.3\textwidth}
				\includegraphics[width=.95\columnwidth,clip]{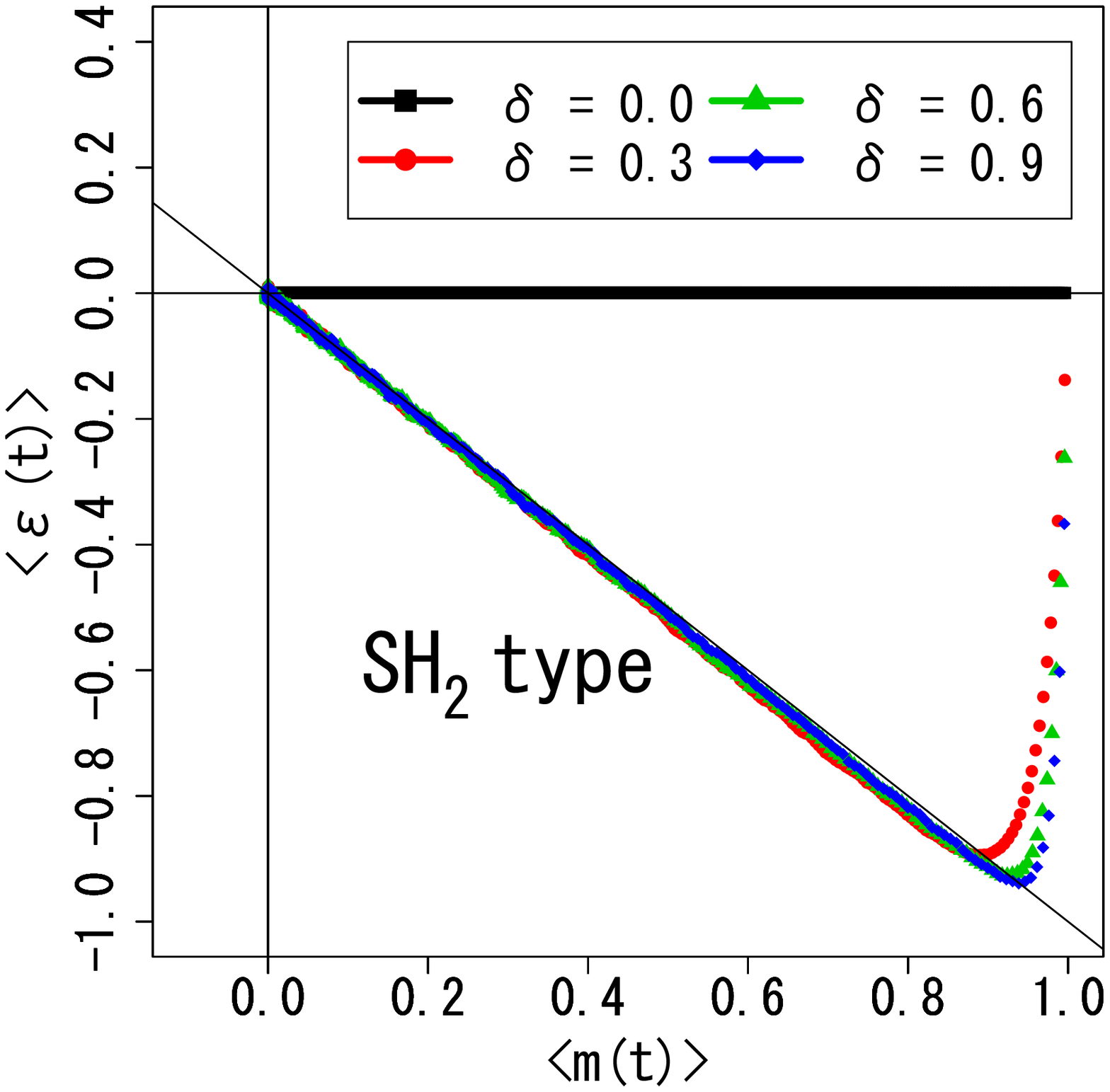}	
				\end{minipage}
			\end{tabular}
		\caption{(Color online) Trajectory of $(\left<m(t)\right>,
		\left<\varepsilon(t)\right>)$ from the initial condition $(1,0)$ 
		to the equilibrium point $(0,0)$ for the SH$_1$ type (left) 
		and the SH$_2$ type (right). 
		The chosen values of parameter in the simulations are 
		$N=2^7$, $\alpha=10^{-2}$, 
		$\gamma=0.6$, $N_{\rm ens}=10^5$, and $M=2.5\times 10^3$.
		Solid lines represent the asymptotic lines 
		near the equilibrium.}
		\label{fig:trajectory1}
		\end{center}
	\end{figure*}

As an illustration of the intrinsic dynamics with SDBC, 
we show the data of trajectory $(\left<m(t)\right>,
\left<\varepsilon(t)\right>)$ in Fig.~\ref{fig:trajectory1} for the 
SH$_1$ and SH$_2$ types. 
In both cases, the slope of the asymptotic line near the equilibrium 
observed in Fig.~\ref{fig:trajectory1} coincides with the theoretical 
prediction discussed in \S\ref{3.4}. 
A deviation from the straight line is found far from the equilibrium point. 
This is due to the effect of high-order correlation, 
approximated to Eq.~(\ref{eq:assumption}) in our analysis, 
for the SH$_1$ type and due to the finite size effect for the SH$_2$ 
type, which disappears in the thermodynamical limit.

\section{Summary and Discussion}

We have studied dynamics of the one-dimensional Ising model 
in a class of the irreversible Markov chain, where 
the SDBC, instead of DBC in the reversible Markov chain, 
ensures the existence of the stationary distribution in a long-time limit.
In particular, the relaxation time of the magnetization density and its
autocorrelation function have been discussed for three different transition
probabilities satisfying SDBC, called the SH$_1$, SH$_2$ and
TCV types, in which the parameter $\delta$ controls
the deviation from DBC. 
In the case of SH$_1$ and SH$_2$ types, 
we have obtained theoretical results of the dynamical behavior 
of the magnetization density and have revealed that 
the relaxation time and the autocorrelation time are 
always reduced for non-zero parameter $\delta$. 
Furthermore, we have shown that the SH$_2$ type transition probability 
changes the dynamical critical exponent while the SH$_1$ type does not.
	\begin{figure}[t]
		\begin{center}
		\includegraphics[width=.6\columnwidth,clip]{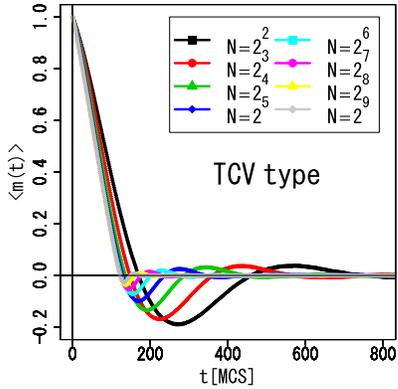}
		\end{center}
	\caption{(Color online) Time evolution of
	the magnetization density for several sizes $N$ 
	with the TCV transition probability. 
	The chosen values of parameter in the simulations are 
	$\alpha=0.05$, $\gamma=0.6$, $\delta=0.9$ and $N_{\rm ens}=10^5$.}
	\label{fig:N_dependence}
	\end{figure}

Some arguments in the previous sections can be made for
more general Ising models in high dimensions. Suppose that the Hamiltonian 
of the Ising system is given as  
	\begin{equation}
	\mathcal{H}(\bmath{\sigma})=
	-\sum_{j<k}J_{jk}\sigma_j\sigma_k-\sum_j H_j\sigma_j, 
	\end{equation}
where $J_{jk}$ denotes an interaction between $\sigma_j$ and $\sigma_k$ and
$H_j$ is a local magnetic field acting on $\sigma_j$. 
The transition probability $w_j(\bmath{\sigma},\varepsilon)$ for the spin
flip discussed here is given as 
	\begin{equation}
	w_j(\bmath{\sigma},\varepsilon)=\frac12\alpha
	\left(1-\sigma_j{\rm tanh}\beta
	 E_j\right)\left(1-\delta\varepsilon\sigma_j\right), 
	\end{equation}
where $E_j=\sum_kJ_{jk}\sigma_k+H_j$ is a local field on $j$-th site,
including the one-dimensional model as $J_{jk}=J\delta_{k,j+1}$. 
This satisfies SDBC and is reduced to the ordinary Glauber transition
probability when $\delta=0$, which has been studied.~\cite{Kubo} 
If the transition probability of the ${\rm SH_1}$ type 
is used for the $\varepsilon$ flip, 
the magnetization density obeys 
	\begin{equation}
	\frac1{\alpha(1+\delta^2)}\frac{d}{dt}\left<m(t)\right>=
	-\left<m(t)\right>
	+\frac1N\sum_j\left<{\rm tanh}\beta E_j\right>, 
	\label{eq:GM}
	\end{equation}
for large $N$.
	\begin{figure}[t]	
		\begin{center}
		\includegraphics[width=.6\columnwidth,clip]{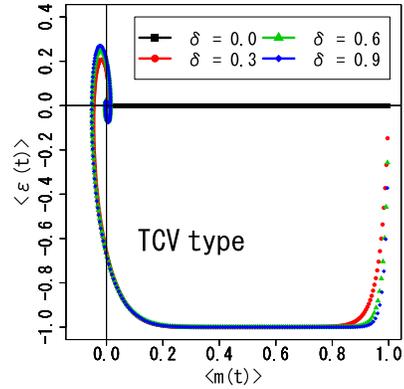}
		\end{center}
	\caption{(Color online) Trajectory of $(\left<m(t)\right>,
	\left<\varepsilon(t)\right>)$ from the initial condition $(1,0)$ 
	to the equilibrium point $(0,0)$ for the TCV type. 
	The chosen values of parameter in the simulations are 
	$N=2^7$, $\alpha=10^{-2}$, 
	$\gamma=0.6$, $N_{\rm ens}=10^5$, and $M=2.5\times 10^3$.}
	\label{fig:trajectory2}
	\end{figure}	
This indicates that the SH$_1$ type yields only the change of time
constant from $\alpha$ in DBC to $\alpha(1+\delta^2)$ in SDBC. 
Hence, the relaxation time is reduced up to the factor $1+\delta^2$ from
DBC, independent of temperature, even when the system exhibits a phase
transition at finite temperature in more than two dimensions. 
Although it is not certain if this argument is valid for other
observables, this argument suggests the existence of a class of 
transition probabilities in the irreversible Markov chain 
which yields a finite gain, temperature independent, 
in the relaxation time
compared to that of the corresponding transition probability with DBC.
	
Finally, let us discuss the eigenvalue in the irreversible Markov chain. 
Although all eigenvalues of a Markov chain with DBC are real 
in general, it is considered that whether an eigenvalue is real 
or complex depends on how to choose transition probabilities 
if the Markov chain does not satisfy DBC.
In the method of SDBC which we consider in this paper, 
several choices of a transition probability 
$w_j(\bmath{\sigma},\varepsilon)$ satisfying SDBC 
in Eq.~(\ref{eq:SDBC}) are possible in general. 
Even if $w_j(\bmath{\sigma},\varepsilon)$ is fixed, 
we can choose some transition probabilities 
$\lambda(\bmath{\sigma},\varepsilon)$ satisfying the condition 
in Eq.~(\ref{eq:lambda}), such as the SH$_1$, SH$_2$, and TCV types. 
As discussed in \S 3, it turns out that the 
eigenvalues which affect to the dynamical behavior of 
the magnetization density are all real in the case of 
the SH$_1$ type. 
However, numerical simulations for the one-dimensional Ising model 
with other transition probabilities show different dynamical 
behavior of the magnetization density. 
For instance, in the case of SH$_2$ type,  
the magnetization density converges exponentially in time 
observed in the middle panel of Fig.~\ref{fig:irN128_r6_m}. 
On the other hand, the existence of complex eigenmode in the 
relaxation dynamics is 
strongly implied from numerical simulations in \S\ref{simulations} 
for the TCV type.
Figure \ref{fig:N_dependence} shows time evolution of the magnetization 
density for several system sizes $N$, indicating that the complex 
eigenvalues clearly depend on $N$ and the imaginary part decreases
with $N$.

Unfortunately little is known about the dynamics in the TCV type analytically.
According to Turitsyn et al.~\cite{Turitsyn},  
a change in the dynamical critical exponent is induced 
in the mean-field Ising model with the use of the TCV type
In fact, as seen in \S 4, the transition probability of the TCV type 
provides the largest reduction of the relaxation time.
This may be explained in the dynamical trajectory shown in 
Fig.~\ref{fig:trajectory2}.
The asymptotic slope, which is not derived analytically, 
is almost vertical, quite far from the horizontal axis 
in the case with DBC.
Therefore, further theoretical studies for the MCMC methods 
without DBC are of prime importance to clarify the 
mechanism which induces the change of dynamics.

\section*{Acknowledgments}
We would like to thank Shiro Ikeda for continuous discussions. 
This research was supported by a Grants-in-Aid for Scientific Research 
from the MEXT, 
Japan, No. 22340109.

\end{document}